\documentclass[letterpaper,journal]{IEEEtran}
\ifdefined\pdfobjcompresslevel
\fi
\usepackage{amsmath,amsfonts,amssymb}
\usepackage{algorithmic}
\usepackage{algorithm}
\usepackage{array}
\usepackage{tabularx}
\usepackage[caption=false,font=normalsize,labelfont=sf,textfont=sf]{subfig}
\usepackage{textcomp}
\usepackage{stfloats}
\usepackage{placeins}
\usepackage{balance}
\usepackage{url}
\usepackage{verbatim}
\usepackage{graphicx}
\usepackage{booktabs}
\usepackage{multirow}
\usepackage{threeparttable}
\usepackage{makecell}
\usepackage{xcolor}
\usepackage{cite}
\usepackage{hyperref}
\hypersetup{hidelinks}
\usepackage{orcidlink}
\newcommand{\TblFont}{\footnotesize}

\newcommand{\TblColSep}{3pt}
\newcommand{\TblStretch}{1.0}
\newcolumntype{A}{>{\centering\arraybackslash}X}
\newcolumntype{E}{>{\centering\arraybackslash}X}
\newcommand{\MR}[2]{\multirow{#1}{*}{#2}}
\newcommand{\eqn}[1]{equation~\eqref{#1}}

\newcommand{\MeanStd}[2]{\mbox{$#1\!\pm\!#2$}}
\newcommand{\BoldMeanStd}[2]{\mbox{$\mathbf{#1\!\pm\!#2}$}}
\newcommand{\MeanStdPct}[3]{\mbox{$#1\!\pm\!#2$}}
\newcommand{\BoldMeanStdPct}[3]{\mbox{$\mathbf{#1\!\pm\!#2}$}}
\begin{document}

% \title{DPD-NeuralEngine~v2: A 12-nm FinFET 1.28-GSa/s Instantaneous-Power-Gated Depthwise Temporal Convolutional Network Accelerator for Wideband Digital Predistortion}
\title{StreamingDPD: A 1.28-GSa/s Instantaneous-Power-Gated Depthwise TCN Accelerator for Real-Time 320-MHz Digital Predistortion}

\author{Yizhuo~Wu\orcidlink{0009-0009-5087-7349}, 
% \IEEEmembership{Graduate Student Member,~IEEE},
Xinling~Wang,
Chang~Gao\textsuperscript{*}\orcidlink{0000-0002-3284-4078}
% \IEEEmembership{Member,~IEEE}
\thanks{\textsuperscript{*}Corresponding author: Chang Gao (e-mail: chang.gao@tudelft.nl).}% <-this % stops a space
\thanks{Yizhuo Wu, Xinling Wang, and Chang Gao are with the Department of Microelectronics, Delft University of Technology, The Netherlands.}}

\maketitle

\begin{abstract}
State feedback between successive samples limits the streaming throughput of recurrent neural networks for digital predistortion (DPD). We present \texttt{StreamingDPD}, an accelerator that processes 320-MHz signals at 1.28\,GSa/s. Its 207-parameter instantaneous-power-gated depthwise temporal convolutional network (IPG-DWTCN) uses $|x[n]|^2$ to gate feature channels and causal dilated depthwise convolutions to model power amplifier memory. The network maps to a spatially unrolled fixed-point pipeline with shift-based quadratic activations and layer-wise mixed precision, sustaining one complex sample per cycle per lane. Its algorithmic cost is 554 floating-point operations per complex sample, 30.1\% fewer than the recurrent baseline with the best linearization performance. In RF playback measurements on three RF devices driven by 40- to 320-MHz Wi-Fi~6/7 and 5G~NR waveforms, IPG-DWTCN achieves the lowest mean adjacent-channel power ratio (ACPR) and ranks first or ties for first in mean error vector magnitude (EVM) against five neural baselines of comparable size. Four-lane FPGA hardware-in-the-loop tests on a 320-MHz 4096-QAM Wi-Fi~7 waveform achieve $-45.82$\,dBc ACPR and $-42.94$\,dB EVM at 1.28\,GSa/s. At the same sample rate, post-layout evaluation of a single-lane 12-nm FinFET ASIC yields a 0.10-mm$^2$ core, a 52-cycle (40.63\,ns) latency, and an estimated typical-corner power of 142.21\,mW (111.10\,pJ/sample). Relative to the prior 22-nm recurrent DPD ASIC, the design provides 5.1~times the throughput at about one-seventh the energy per sample.
\end{abstract}


\IEEEpeerreviewmaketitle

%======================================================================
\section{Introduction}
\label{sec:introduction}
%======================================================================
\IEEEPARstart{T}{he} rapid growth of wireless traffic is driving 5G-advanced, emerging 6G, and Wi-Fi~7 transmitters toward multi-hundred-megahertz baseband bandwidths ($f_{\mathrm{BB}}$) and constellations up to 4096-QAM. To maintain radio-frequency (RF) efficiency, power amplifiers (PAs) operate near saturation, where nonlinearities and broadband memory degrade in-band fidelity and regrow adjacent-channel emissions. Digital predistortion (DPD) applies the inverse PA characteristic in baseband, but widening $f_{\mathrm{BB}}$ pushes continuous in-phase/quadrature (I/Q) processing from hundreds of megasamples per second (MSa/s) into the gigasample-per-second (GSa/s) regime, such as 1.28\,GSa/s for a 320-MHz Wi-Fi~7 channel at $4\times$ oversampling, under stringent silicon-area, deterministic-latency, and sub-watt power budgets~\cite{Li2025ISCAS,10994208,11449061}.

\begin{figure}[!t]
\centering
\includegraphics[width=0.95\columnwidth]{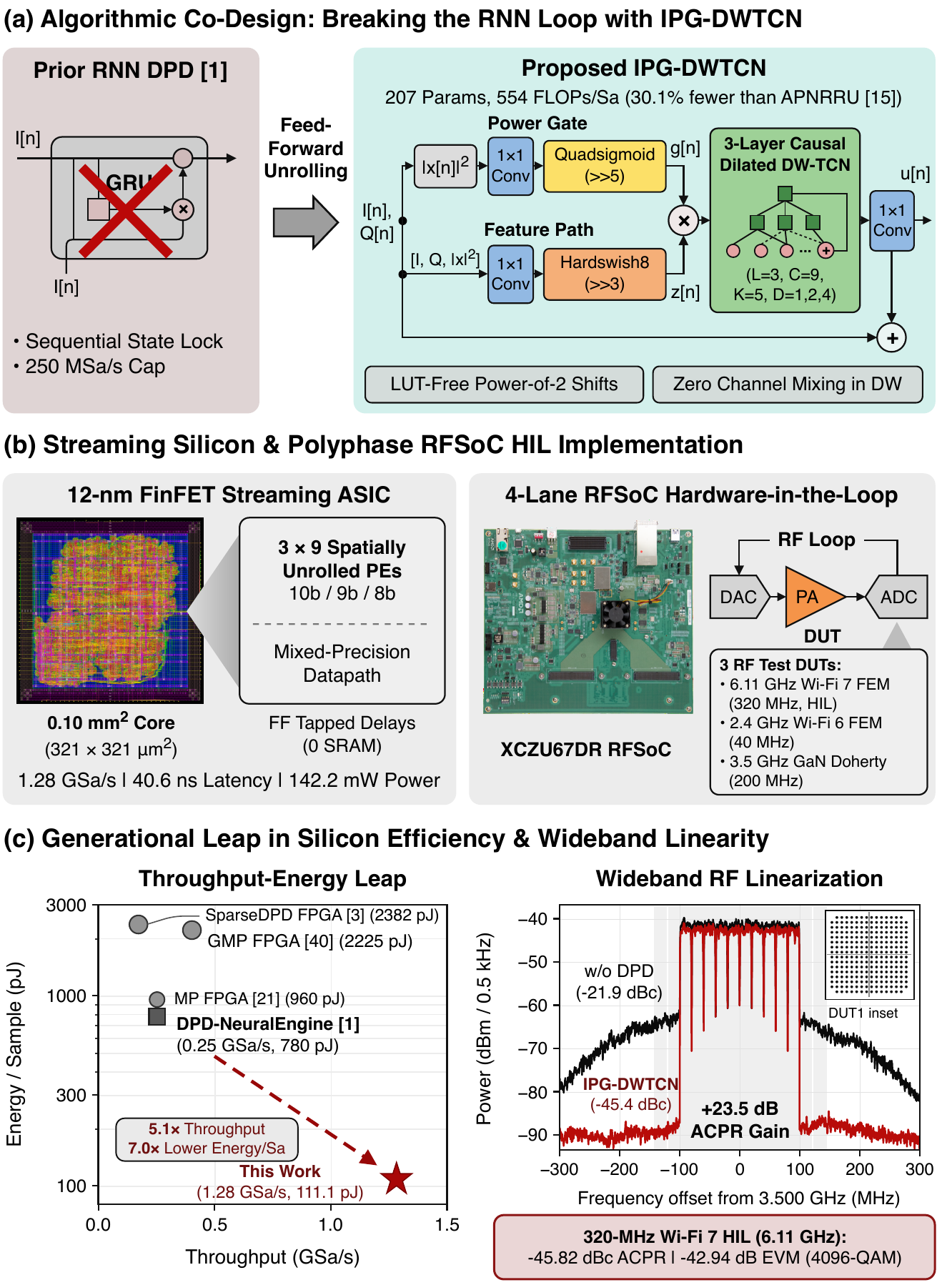}
\caption{Overview of \texttt{StreamingDPD}: (a)~recurrence-free IPG-DWTCN with instantaneous-power gating and shift-based activations, (b)~12-nm FinFET ASIC layout and four-lane RFSoC HIL setup for DUT1, and (c)~throughput-energy comparison, DUT3 spectrum, and DUT1 HIL results with constellation inset.}
\label{fig:overview}
\end{figure}

Algorithmically, industrial DPD relies on Volterra models such as the generalized memory polynomial (GMP)~\cite{GMP}, whose cross-memory expansion grows costly across wider bandwidths. Neural networks offer higher parameter efficiency: time-delay neural networks (TDNNs) capture short-term memory~\cite{Rawat2010}, whereas recurrent architectures model sequence dynamics via long short-term memory (LSTM)~\cite{Hochreiter1997LSTM}, gated recurrent units (GRUs)~\cite{Cho2014GRU}, Just Another NETwork (JANET)~\cite{van2018unreasonable}, and domain-structured variants using vector decomposition, phase gating, decomposed vector rotation (DVR-JANET), phase normalization, and block-oriented structures (BO-JANET)~\cite{VDLSTM,PGJANET,DVRJANET,PNRNN,PhaseNormalizedNN2023,BOJANET,APNRRU}. These structured models surpass GMP at compact sizes, improving adjacent-channel power ratio (ACPR) by $>3$\,dB at 100\,MHz~\cite{DVRJANET}, gaining 4.0\,dB to reach below $-50$\,dBc~\cite{PhaseNormalizedNN2023}, and generalizing across 2.40- and 3.55-GHz PAs~\cite{BOJANET}. In parallel, convolutional neural networks (CNNs)~\cite{Hu2022RVTDCNN} and dilated temporal convolutional networks (TCNs), including our noncausal TCN-DPD~\cite{Duan2025TCN-DPD}, expand receptive fields without recurrence. Yet a fundamental tension persists for GSa/s silicon: recurrent gates capture envelope-dependent nonlinearity compactly, but state feedback ($\mathbf{h}[n]$ from $\mathbf{h}[n-1]$) serializes sample execution; plain CNNs and TCNs remove recurrence, but lack explicit envelope modulation and rely on costly cross-channel mixing and transcendental activations.

This tension creates a throughput-energy wall in DPD hardware~\cite{11373642}. Model-based DPDs on field-programmable gate arrays (FPGAs)~\cite{Huang2020,cao2017modified,Cappello2022}, moving-average nested GMP~\cite{Chen2022MANGMP}, parallel delta-sigma DPD~\cite{Othmani2023ParallelDSM}, crosstalk-canceling multiple-input multiple-output (MIMO) DPD~\cite{Wu2024DCCDPD}, and commercial pruned-GMP transceivers like the ARM-assisted AD9375~\cite{AD9375,AD9375UG992} remain confined to analytical structures. Among neural designs, graphics processing units (GPUs) reach about 1\,GSa/s in batched kernel timing tests~\cite{10994208}, while SparseDPD reports 170\,MSa/s with an estimated 2.38\,nJ/sample on FPGA~\cite{11449061}. Compact recurrent hardware, including low-latency residual GRUs~\cite{11278888} and our 22-nm GRU application-specific integrated circuit (ASIC), \texttt{DPD-NeuralEngine}, demonstrates silicon viability, yet state recurrence restricts to 250\,MSa/s and 780\,pJ/sample, over $5\times$ short of 1.28\,GSa/s~\cite{Li2025ISCAS}. Edge recurrent~\cite{gao2020edgedrnn,gao2024spartus} and general deep-learning processors~\cite{Molendijk2023,Lin2020} likewise target lower-rate workloads rather than continuous one-sample/cycle GSa/s I/Q streaming.

To bridge this gap, we present \texttt{StreamingDPD} (Fig.~\ref{fig:overview}), co-designing a recurrence-free 207-parameter instantaneous-power-gated depthwise TCN (IPG-DWTCN) with an unrolled fixed-point streaming architecture at an initiation interval (II) of one cycle, verified across physical devices under test (DUTs), bit-accurate register-transfer-level (RTL) simulation, four-lane FPGA hardware-in-the-loop (HIL) testing, and 12-nm FinFET post-layout evaluation. The main contributions are as follows:
\begin{enumerate}
    \item \textbf{Recurrence-Free IPG-DWTCN Architecture:} IPG-DWTCN modulates features via an $|x[n]|^2$-driven gate, captures PA memory through channel-separated causal dilated depthwise convolutions, and employs shift-friendly quadratic activations (\texttt{Quadsigmoid} and \texttt{Hardswish8}). With 207 parameters and 554 floating-point operations (FLOPs) per complex I/Q sample, the three-layer, nine-channel (L3/C9) configuration reduces computation by 30.1\% over the closest-accuracy baseline~\cite{APNRRU}.
    \item \textbf{Multi-Technology Wideband RF Benchmarking:} Offline-waveform RF measurements against five neural baselines near a 200-parameter budget across a 6.11-GHz Wi-Fi~7 front-end module driven by a 320-MHz extremely high throughput (EHT) 4096-QAM channel, a 2.40-GHz Wi-Fi~6 front-end module (40-MHz 1024-QAM), and a 35.20-dBm 3.50-GHz gallium nitride (GaN) Doherty PA driven by a ten-carrier 200-MHz 5G New Radio (NR) signal achieve the best mean ACPR ($-45.90$, $-46.84$, and $-45.39$\,dBc) and best or tied error vector magnitude (EVM) ($-42.91$, $-43.94$, and $-39.85$\,dB), delivering up to 23.47-dB ACPR and 19.88-dB EVM gain.
    \item \textbf{1.28-GSa/s FPGA HIL and 12-nm FinFET ASIC Realization:} A RAM-free post-route four-lane polyphase core on an XCZU67DR radio-frequency system-on-chip (RFSoC) with 12-bit parameter storage and 16-bit I/Q (W12A16 interface) achieves $-45.82$\,dBc ACPR and $-42.94$\,dB EVM on DUT1 at 320\,MHz and 1.28\,GSa/s (Table~\ref{tab:soa}). Single-lane 12-nm post-layout estimates yield a 0.10-mm$^2$ core, 1.28\,GSa/s at II~$=1$ (52 cycles, 40.63\,ns), and 142.21\,mW typical-typical (TT) power (111.10\,pJ/sample), providing $5.1\times$ higher throughput and $7.0\times$ lower energy/sample than~\cite{Li2025ISCAS}.
\end{enumerate}

%======================================================================
\section{The IPG-DWTCN Algorithm}
\label{sec:algorithm}
\begin{figure*}[t]
\centering
\includegraphics[width=\linewidth]{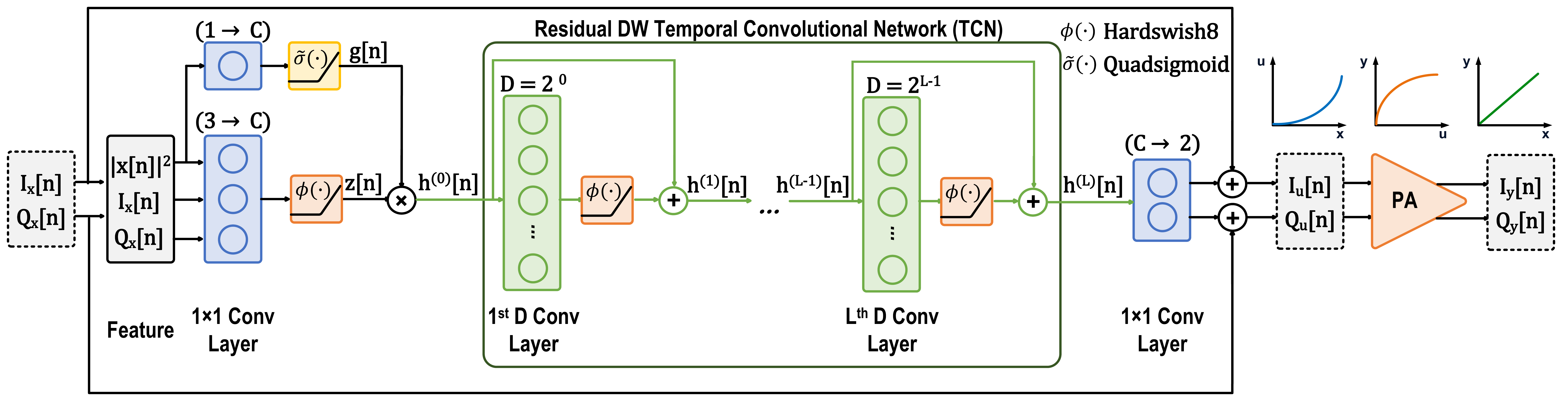}
\caption{Architecture of the proposed IPG-DWTCN DPD model. The final hidden feature is $\mathbf{h}^{(L)}[n]$, with $L=3$ and $C=9$ in the selected model.}
\label{fig:pgtcn}
\end{figure*}

%======================================================================
Fig.~\ref{fig:pgtcn} illustrates IPG-DWTCN. Extending our noncausal TCN-DPD~\cite{Duan2025TCN-DPD} to GSa/s streaming, IPG-DWTCN enforces strict causality and pairs an instantaneous-power gate with $L$ channel-separated dilated depthwise-convolution blocks across $C$ channels ($L=3$, $C=9$ in the selected L3/C9 model), shift-based quadratic activations, and an I/Q skip-connected output projection. The gate modulates channels via $|x[n]|^2=I_x^2[n]+Q_x^2[n]$, while the causal stack models PA memory without recurrence or look-ahead. Below, $x[n]=I_x[n]+jQ_x[n]$, $u[n]=I_u[n]+jQ_u[n]$, and $y[n]=I_y[n]+jQ_y[n]$ denote the DPD input, DPD output, and PA output, respectively, and $\odot$ denotes element-wise multiplication.

% These two figures are declared early so that both retain top-of-page
% placement while the wide accelerator view can appear one page earlier.

\subsection{Instantaneous-Power Feature Gating}
At sample $n$, the front end forms the three-element input vector
\begin{equation}
\mathbf{f}[n]=\big[\,I_x[n],\; Q_x[n],\; |x[n]|^2\,\big]^{\!\top}\in\mathbb{R}^{3}.
\label{eq:feat}
\end{equation}
A $1\times1$ affine projection ($\mathbf{W}_{\mathrm{in}}\in\mathbb{R}^{C\times 3}$, $\mathbf{b}_{\mathrm{in}}\in\mathbb{R}^{C}$) maps $\mathbf{f}[n]$ to $C$ channels through the element-wise \texttt{Hardswish8} activation $\phi(\cdot)$ defined in \eqn{eq:hsw8}:
\begin{equation}
\mathbf{z}[n]=\phi\!\big(\mathbf{W}_{\mathrm{in}}\,\mathbf{f}[n]+\mathbf{b}_{\mathrm{in}}\big).
\label{eq:convin}
\end{equation}
In parallel, a second $1\times1$ branch ($\mathbf{w}_{\mathrm{pg}},\mathbf{b}_{\mathrm{pg}}\in\mathbb{R}^{C}$) maps $|x[n]|^2$ to $C$ gates in $[0,1]$ via the element-wise \texttt{Quadsigmoid} activation $\tilde{\sigma}(\cdot)$ defined in \eqn{eq:pqs}:
\begin{equation}
\mathbf{g}[n]=\tilde{\sigma}\!\big(\mathbf{w}_{\mathrm{pg}}\,|x[n]|^2+\mathbf{b}_{\mathrm{pg}}\big).
\label{eq:gate}
\end{equation}
Channel-wise gating yields the temporal-stack input:
\begin{equation}
\mathbf{h}^{(0)}[n]=\mathbf{z}[n]\odot\mathbf{g}[n].
\label{eq:gating}
\end{equation}
Through $(\mathbf{w}_{\mathrm{pg}},\mathbf{b}_{\mathrm{pg}})$, channels specialize across linear, compression, and saturation drive regimes without recurrent gates, while reusing $|x[n]|^2$ from $\mathbf{f}[n]$ avoids the square-root or coordinate-rotation logic needed by amplitude- or phase-extraction front ends.

\subsection{Residual Dilated Depthwise Convolution Stack}
\label{sec:stack}
PA memory is modeled by $L$ cascaded residual dilated depthwise (DW) blocks holding most model parameters. Block $l\in\{1,\dots,L\}$ applies a bias-free causal DW convolution of kernel size $K$ and dilation $D_l=2^{l-1}$ per channel:
\begin{equation}
d_c^{(l)}[n]=\sum_{k=0}^{K-1} w^{(l)}_{c,k}\,h^{(l-1)}_c\!\big[n-kD_l\big],\qquad c=1,\dots,C,
\label{eq:dwconv}
\end{equation}
where $h_c^{(l-1)}[n]$ is channel $c$ of $\mathbf{h}^{(l-1)}[n]$, $w_{c,k}^{(l)}$ is its $k$th tap weight, and $h_c^{(l-1)}[m]=0$ for $m<0$ via causal left zero-padding in training and zeroed tap histories at stream start. Stacking $\mathbf{d}^{(l)}[n]=\big[d_1^{(l)}[n],\dots,d_C^{(l)}[n]\big]^{\!\top}\in\mathbb{R}^{C}$, each block computes
\begin{equation}
\mathbf{h}^{(l)}[n]=\mathbf{h}^{(l-1)}[n]+\phi\!\big(\mathbf{d}^{(l)}[n]\big).
\label{eq:resblock}
\end{equation}
Looking back $(K-1)D_l$ samples per block without look-ahead or channel mixing yields a causal receptive field of $1+\sum_{l=1}^{L}(K-1)D_l=1+(K-1)(2^L-1)$ samples, allowing continuous execution as samples arrive. For the selected configuration ($L=3$, $K=5$, $D_l\in\{1,2,4\}$), this window spans 29 samples, or 28 sample intervals (21.88\,ns at 1.28\,GSa/s).

A bias-free $1\times1$ projection $\mathbf{W}_{\mathrm{out}}\in\mathbb{R}^{2\times C}$ maps $\mathbf{h}^{(L)}[n]$ to an additive I/Q correction around the linear input path:
\begin{equation}
\begin{bmatrix}I_u[n]\\Q_u[n]\end{bmatrix}=\mathbf{W}_{\mathrm{out}}\,\mathbf{h}^{(L)}[n]+\begin{bmatrix}I_x[n]\\Q_x[n]\end{bmatrix}.
\label{eq:out}
\end{equation}
Channel mixing is thus confined to the $1\times1$ projections. Summing $3C+C$ (input), $C+C$ (gate), $LKC$ (DW stack), and $2C$ (output) gives $(8+LK)C$ parameters ($(6+LK)C$ without gating), totaling 207 parameters for $L=3$, $K=5$, and $C=9$.

\subsection{Hardware-Friendly Activation Functions}
\label{sec:act}
\begin{figure}[!t]
    \centering
    \includegraphics[width=\linewidth]{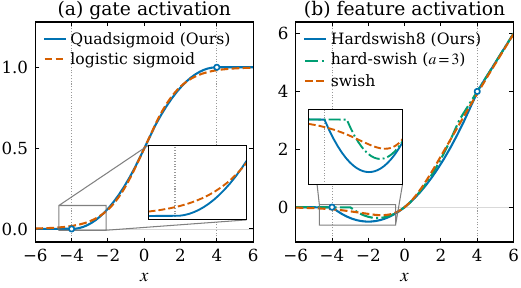}
    \caption{Activations adopted in IPG-DWTCN: (a)~gate activation \texttt{Quadsigmoid} versus logistic sigmoid and (b)~feature activation \texttt{Hardswish8} versus MobileNetV3 hard-swish ($a=3$) and swish. Insets magnify local approximation differences.}
    \label{fig:act}
\end{figure}
\begin{figure*}[!t]
    \centering
    \includegraphics[width=\linewidth]{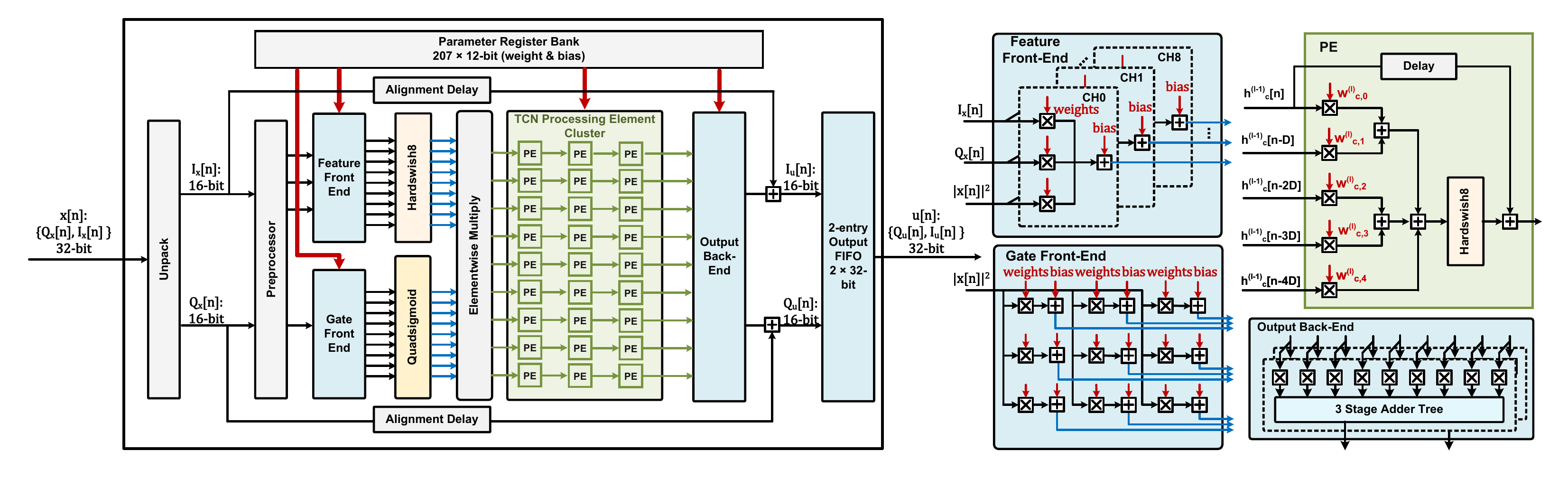}
    \caption{Microarchitecture of the single-lane IPG-DWTCN accelerator: streaming datapath (left) and front-end, DW channel PE, and output operators (right).}
    \label{fig:dwblock}
\end{figure*}

Logistic sigmoid $\sigma(x)=(1+e^{-x})^{-1}$ and swish $x\sigma(x)$ require exponentials and divisions that demand look-up tables (LUTs) or dividers at GSa/s rates. Following piecewise-quadratic sigmoid generators~\cite{Zhang1996Sigmoid}, we adopt a $C^1$-continuous two-segment quadratic approximation matching $\sigma(0)=0.5$ and $\sigma'(0)=0.25$ with $\pm4$ saturation, termed \texttt{Quadsigmoid}, for the power gate:
\begin{equation}
\tilde{\sigma}(x)=
\begin{cases}
0, & x\le -4,\\[2pt]
\dfrac{(x+4)^2}{32}, & -4<x<0,\\[6pt]
1-\dfrac{(x-4)^2}{32}, & 0\le x<4,\\[6pt]
1, & x\ge 4.
\end{cases}
\label{eq:pqs}
\end{equation}

For the feature and DW paths, MobileNetV3 hard-swish (denoted HardSwish), $x\operatorname{ReLU6}(x+3)/6$ with $\operatorname{ReLU6}(v)=\min(\max(v,0),6)$~\cite{Howard2019MobileNetV3}, removes the exponential of swish but retains division by $6$. Shifting its breakpoints from $\pm3$ to $\pm4$ and replacing $1/6$ with $1/8$ gives \texttt{Hardswish8}:
\begin{equation}
\phi(x)=\mathrm{Hardswish8}(x)=
\begin{cases}
0, & x\le -4,\\[2pt]
x\cdot\dfrac{x+4}{8}, & -4<x<4,\\[6pt]
x, & x\ge 4,
\end{cases}
\label{eq:hsw8}
\end{equation}
preserving $\phi(0)=0$, $\phi'(0)=0.5$, and breakpoint continuity. Since the active quadratic numerator in \eqn{eq:pqs} reduces to $(4-|x|)^2$ across $(-4,4)$ and the divisors $32$ and $8$ become 5- and 3-bit right shifts, both $\tilde{\sigma}(\cdot)$ and $\phi(\cdot)$ require only one multiplication in a shared add-multiply-compare-shift datapath without a LUT or general divider. Fig.~\ref{fig:act}(a) and~(b) compare both activations with their conventional counterparts; maximum absolute deviations from logistic sigmoid and swish are $0.02$ and $0.28$, respectively.

%======================================================================
\section{Hardware Architecture}
\label{sec:hw}
%======================================================================
Fig.~\ref{fig:dwblock} shows the single-lane ($P=1$) streaming datapath (left) and its constituent operators (right). Replacing state recurrence with finite causal tap histories allows the gated front end, $L=3$ temporal stages, and output back end to be spatially unrolled across a 52-cycle pipeline (16 front-end, $3\times10$ temporal, and 6 back-end cycles) fed by a 207-entry parameter bank. Once primed, the core sustains II~$=1$ ($f_s=f_{\mathrm{clk}}$) and extends directly to $P$-lane polyphase execution ($f_s=Pf_{\mathrm{clk}}$).

\subsection{Architecture Overview and Dataflow}
\label{sec:datapath}
The AXI-Stream-style interface transfers packed 32-bit $\{Q,I\}$ words under \texttt{valid}/\texttt{ready} handshaking. Unpacked $I_x[n]$ and $Q_x[n]$ enter a 52-stage alignment delay line and, in parallel, the four-cycle preprocessor computing $|x[n]|^2=I_x^2[n]+Q_x^2[n]$. Tapping the delay line at cycle~4 aligns $I_x[n],Q_x[n]$ with $|x[n]|^2$ to form $\mathbf{f}[n]$ for the feature front end (\texttt{Hardswish8} output $\mathbf{z}[n]$), while the gate front end maps $|x[n]|^2$ through \texttt{Quadsigmoid} to $\mathbf{g}[n]$. Matched eight-cycle front-end branches synchronize at the element-wise multiplier to form $\mathbf{h}^{(0)}[n]=\mathbf{z}[n]\odot\mathbf{g}[n]$ at cycle~16.

The TCN processing element (PE) cluster cascades $L=3$ stages ($D_l\in\{1,2,4\}$) of $C=9$ channel-independent PEs ($LC=27$ total). Each 10-cycle PE (Fig.~\ref{fig:dwblock}, top right) stores $(K-1)D_l$ past samples in a tapped shift register, multiplies the five causal taps $h_c^{(l-1)}[n-kD_l]$ by $w_{c,k}^{(l)}$, requantizes the sum through \texttt{Hardswish8}, and adds a delay-matched bypass of $h_c^{(l-1)}[n]$. Across all 27 PEs, causal histories total $(4+8+16)\times9=252$ words of 15 bits, requiring only 3780 flip-flops (FFs) (3.69\,kb) with zero SRAM/BRAM or address generation. Channel mixing occurs solely in the $1\times1$ projections. The six-cycle output back end maps $\mathbf{h}^{(3)}[n]$ to an I/Q correction added to tap~52 of the alignment delay line to produce $u[n]$.

Parameters reside in a 207-register bank updated via a stop-drain-program-commit sequence: software deasserts \texttt{valid}, drains in-flight samples, writes the image, and commits, zeroing valid flags and tap histories before restart without adding multiplexers to critical paths. A two-entry first-in, first-out (FIFO) elastic buffer decouples downstream \texttt{ready} backpressure from the high-fanout pipeline stall enable. For lower-clock FPGAs, the datapath scales to $P$-lane polyphase execution ($f_s=Pf_{\mathrm{clk}}$) over a $32P$-bit packed I/Q bus by replicating pointwise paths with shared coefficients and drawing causal DW taps from the current $P$-sample group and a shared $P$-sample/cycle shift history.

\subsection{Layer-Specific Operator Realization}
\label{sec:operators}
All $1\times1$ projections share a parameterized dot-product unit specialized by fan-in. For channel $c$, the gate ($w_{\mathrm{pg},c}|x[n]|^2+b_{\mathrm{pg},c}$) and feature ($w_{c,0}I_x[n]+w_{c,1}Q_x[n]+w_{c,2}|x[n]|^2+b_{\mathrm{in},c}$) front ends execute in four cycles: registered multiplication, two-group partial sum, bias-aligned sum, and requantization. Since $I_x[n],Q_x[n]$ carry 13 fractional bits and $|x[n]|^2$ carries 11, the power product is left-shifted by two bits before accumulation with scale-aligned biases. The bias-free output back end sums $C=9$ weighted channels per I/Q component via the three-stage adder tree in Fig.~\ref{fig:dwblock}, adding one cycle (five total) for GSa/s timing closure. Both \texttt{Quadsigmoid} and \texttt{Hardswish8} use a matched three-cycle pipeline: breakpoint offset ($4-|x|$ or $x+4$), single squaring or multiplication, and right-shift, muxing, and clamping.

Motivated by mixed-precision DPD~\cite{10502240}, post-training quantization (PTQ) maps the frozen 32-bit floating-point (FP32) model to tensor-specific signed two's-complement formats $S(B,F)$ calibrated from coefficient and activation ranges (Table~\ref{tab:fixed-formats}), where $B$-bit integer $m$ represents $m\cdot 2^{-F}$ (allowing $F\ge B$ for small biases such as $S(8,9)$). Software rounds parameters to nearest even in $S(B,F)$ and sign-extends them into uniform 12-bit registers; hardware slices the lower $B\in[8,11]$ bits without runtime rescaling. Multiplying 15-bit stream features by 10-, 9-, and 8-bit weights across DW1-DW3 shrinks the five-tap accumulators to 27, 26, and 25 bits. Accumulators retain full multiply-accumulate (MAC) precision before arithmetic right-shift truncation and signed saturation; residual adders saturate at their common binary point.

\begin{table}[!t]
\centering
\caption{Deployed fixed-point formats for the L3/C9 accelerator.}
\label{tab:fixed-formats}
\scriptsize
\setlength{\tabcolsep}{3.0pt}
\renewcommand{\arraystretch}{1.05}
\begin{tabularx}{\columnwidth}{|>{\raggedright\arraybackslash}p{0.245\columnwidth}|>{\raggedright\arraybackslash}X|}
\hline
\textbf{Stage or quantity} & \textbf{Signed format $S(B,F)$} \\ \hline\hline
Signal interface & I/Q and output: $S(16,13)$; power: $S(16,11)$ \\ \hline
Input projection & Weight/bias: $S(11,10)/S(8,9)$; pre-act./act.: $S(15,11)/S(12,11)$ \\ \hline
Gate branch & Weight/bias: $S(8,6)/S(10,7)$; pre-act./act.: $S(15,10)/S(16,14)$ \\ \hline
Temporal datapath & Gated $\mathbf{h}^{(0)}$/stream $\mathbf{h}^{(l)}$: $S(14,13)/S(15,13)$; DW pre-act.: $S(14,13)$ \\ \hline
DW1/DW2/DW3 weights & $S(10,7)/S(9,7)/S(8,7)$ \\ \hline
Output projection & Weight/correction: $S(10,8)/S(16,15)$ \\ \hline
Parameter storage & 207 signed 12-bit entries \\ \hline
\end{tabularx}
\end{table}

%======================================================================
\section{Experimental Setup}
\label{sec:setup}
%======================================================================
Table~\ref{tab:experimental_setup} summarizes the waveforms and training settings for the three physical DUTs---a Wi-Fi~7 front-end module (DUT1, 320-MHz 4096-QAM), a Wi-Fi~6 front-end module (DUT2, 40-MHz 1024-QAM), and a high-power GaN Doherty PA (DUT3, ten-carrier 200-MHz 5G~NR)---followed by our RF measurement, training, and FPGA/ASIC evaluation setups. DUT1 guides architecture selection and ablations, whereas DUT2 and DUT3 test cross-technology generalization without retuning.
\begin{figure}[t]
    \centering
    \includegraphics[width=\linewidth]{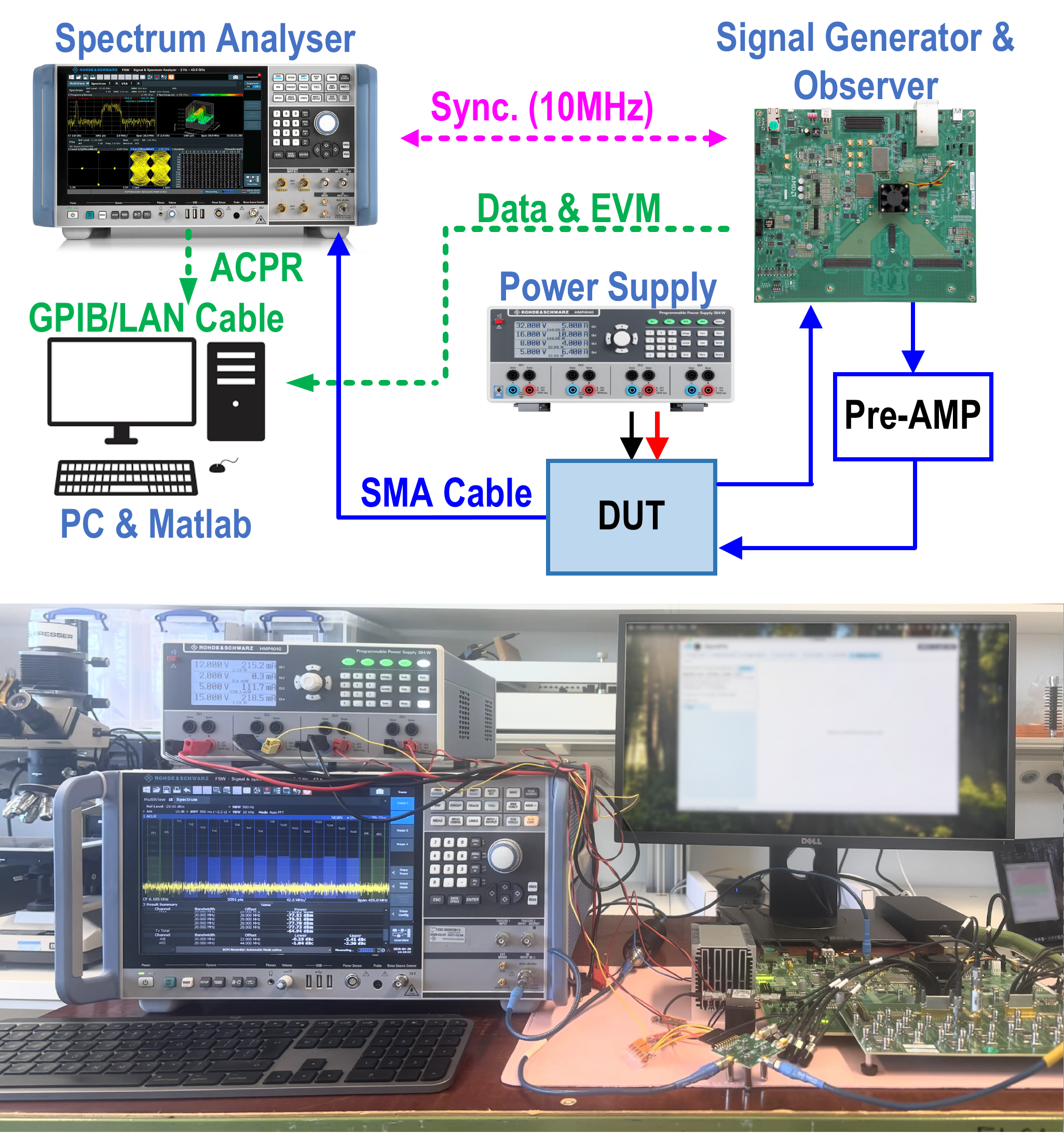}
    \caption{Setup for dataset acquisition and DPD performance measurement.}
    \label{fig:platform}
    \end{figure}

\begin{table}[!t]
\centering
\caption{Experimental setup parameters.}
\label{tab:experimental_setup}
\scriptsize
\setlength{\tabcolsep}{1pt}
\renewcommand{\arraystretch}{1.2}
\newcommand{\dutphoto}[1]{%
  $\vcenter{\vskip 2pt\hbox{#1}\vskip 2pt}$%
}
\begin{tabular*}{\columnwidth}{@{\extracolsep{\fill}}|c|c|c|c|c|}
\hline
\textbf{Category} & \textbf{Parameter} & \textbf{DUT1} & \textbf{DUT2} & \textbf{DUT3} \\ \hline\hline
\multicolumn{2}{|c|}{Photograph} & \dutphoto{\includegraphics[width=40pt]{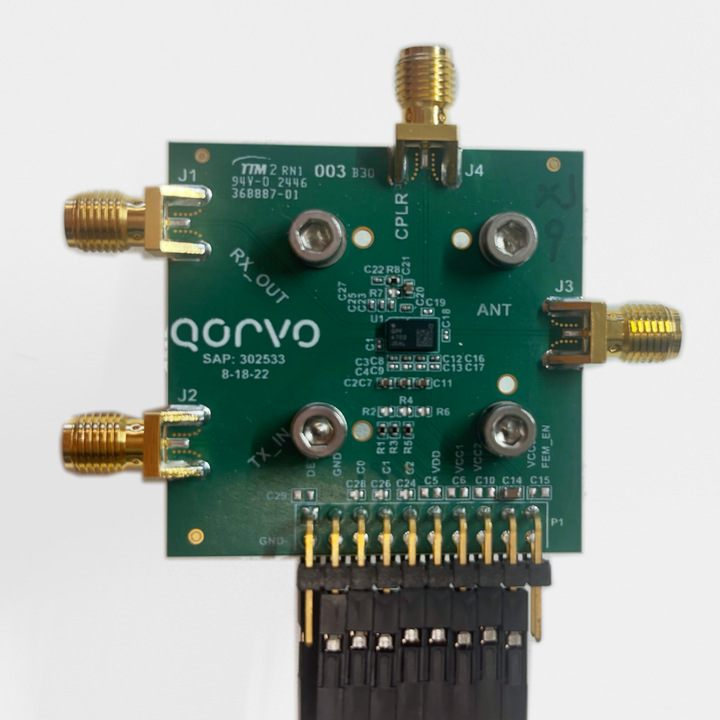}} & \dutphoto{\includegraphics[width=40pt]{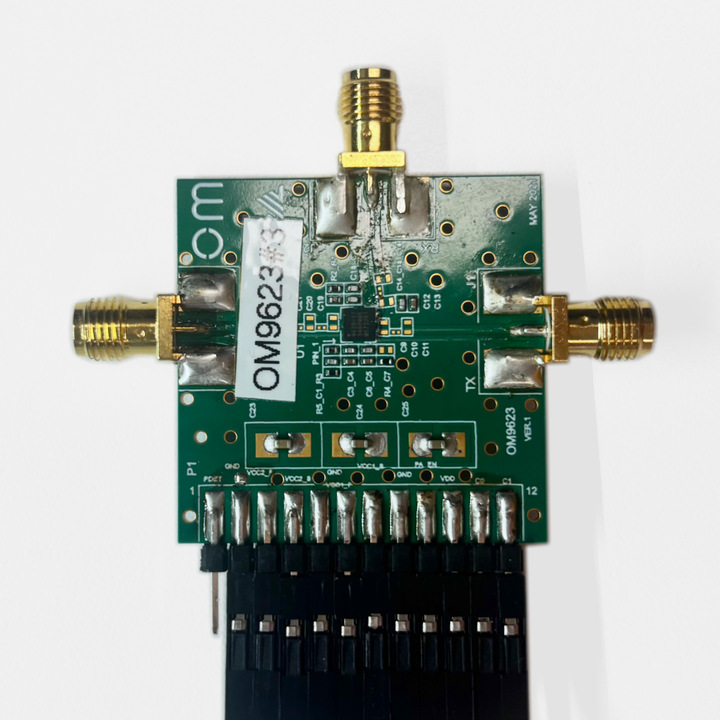}} & \dutphoto{\includegraphics[width=40pt]{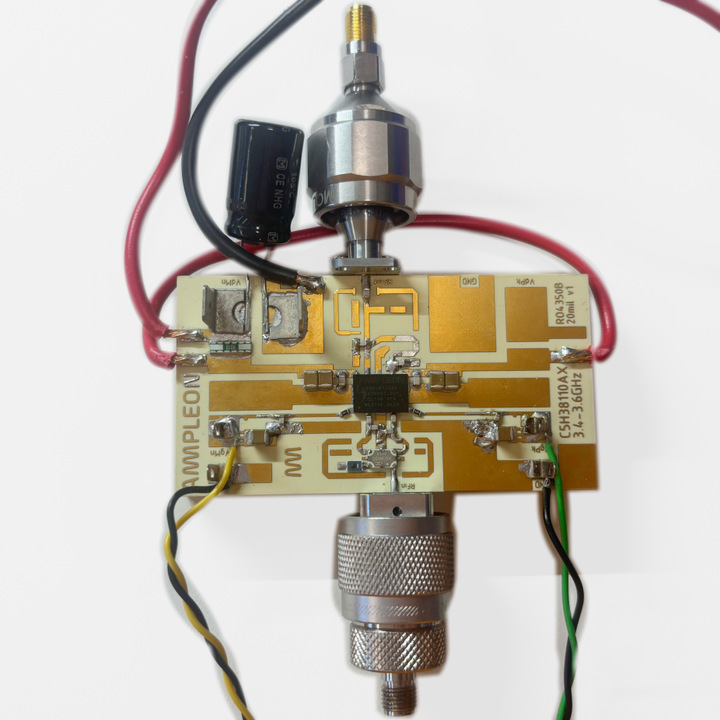}} \\ \hline
\multirow{3}{*}{DUT} & Device & \makecell{Wi-Fi~7\\front-end\\module} & \makecell{Wi-Fi~6\\front-end\\module} & \makecell{3.50-GHz GaN\\Doherty PA} \\ \cline{2-5}
 & RF center & 6.11\,GHz & 2.40\,GHz & 3.50\,GHz \\ \cline{2-5}
 & \makecell{Avg. output power\\(with DPD)} & 23.60\,dBm & 21.80\,dBm & 35.20\,dBm \\ \hline\hline
\multirow{2}{*}{Measurement} & Platform & \multicolumn{3}{c|}{ZCU670 RFSoC} \\ \cline{2-5}
 & Sample rate & 1.28\,GSa/s & 163.84\,MSa/s & 1.20\,GSa/s \\ \hline\hline
\multirow{5}{*}{Test signal} & Waveform & \makecell{EHT-320\,MHz\\MCS13} & \makecell{HE-40\,MHz\\MCS11} & \makecell{5G NR-200\,MHz\\$10\times20$\,MHz} \\ \cline{2-5}
 & Modulation & 4096-QAM & 1024-QAM & 1024-QAM \\ \cline{2-5}
 & Occupied BW & 320\,MHz & 40\,MHz & 200\,MHz \\ \cline{2-5}
 & PAPR & 9.00\,dB & 8.40\,dB & 8.00\,dB \\ \cline{2-5}
 & Train / val / test & \multicolumn{3}{c|}{6:2:2} \\ \hline\hline
\multirow{8}{*}{Training} & PA model & \makecell{GRU, $H=11$\\519 params} & \makecell{GRU, $H=11$\\519 params} & \makecell{GRU, $H=29$\\2931 params} \\ \cline{2-5}
 & PA test NMSE & $-37.66$\,dB & $-40.09$\,dB & $-38.82$\,dB \\ \cline{2-5}
 & Optimizer / init.\ LR & \multicolumn{3}{c|}{AdamW / $5\times10^{-3}$} \\ \cline{2-5}
 & Epochs / batch size & \multicolumn{3}{c|}{100 / 64} \\ \cline{2-5}
 & Frame length / stride & \multicolumn{3}{c|}{200 / 1} \\ \cline{2-5}
 & DPD seeds / selection & \multicolumn{3}{c|}{5 / best validation ACPR} \\ \cline{2-5}
 & Software & \multicolumn{3}{c|}{PyTorch 2.11.0 / CUDA 12.6} \\ \cline{2-5}
 & Hardware & \multicolumn{3}{c|}{NVIDIA RTX 4070 Ti} \\ \hline
\end{tabular*}
\end{table}

\subsection{RF Measurement and Data Splits}
In Fig.~\ref{fig:platform}, a host PC coordinates an AMD ZCU670 RFSoC (baseband generator and observation receiver), pre-amplifier, DC supply, and spectrum analyzer locked to a 10-MHz reference. Baseband orthogonal frequency-division multiplexing (OFDM) signals synthesized in-house match the occupied bandwidth, quadrature amplitude modulation (QAM) order, modulation and coding scheme (MCS), and peak-to-average power ratio (PAPR) of IEEE~802.11be (EHT-320), IEEE~802.11ax (HE-40), and 3GPP 5G~NR ($10\times20$\,MHz) on active data subcarriers without pilot or framing overhead. Captured I/Q pairs undergo fractional-delay alignment and amplitude normalization. Five disjoint random-payload waveforms per DUT are split 3:1:1 (60\%/20\%/20\%, or 6:2:2) into training, validation (model/checkpoint selection), and held-out test sets under fixed MCS and PAPR.

To compare all five baselines, including recurrent models unable to run at 1.28\,GSa/s in real time, across five seeds under identical RF conditions, Tables~\ref{tab:gating-ablation}-\ref{tab:model-comparison} use offline RF playback, whereas real-time 1.28-GSa/s streaming HIL execution is validated in Section~\ref{sec:hwresults}. On DUT1, all five IPG-DWTCN seeds use bit-accurate mixed precision (Table~\ref{tab:fixed-formats}, calibrated on seed-0 validation and frozen); all DUT1 baselines and all DUT2/DUT3 runs use FP32 inference before 16-bit I/Q export. Input drive is adjusted to match the average output power of each DUT with and without DPD (Table~\ref{tab:experimental_setup}).

Side-$s$ ACPR ($s\in\{\mathrm{L},\mathrm{R}\}$) is computed from integrated linear powers as $10\log_{10}(P_{\mathrm{adj},s}/P_{\mathrm{ref},s})$ and averaged in decibels over both sides. DUT1 and DUT2 integrate $P_{\mathrm{ref},s}$ and $P_{\mathrm{adj},s}$ over 320- and 40-MHz bandwidths (adjacent centers at $\pm352$ and $\pm44$\,MHz); ten-carrier DUT3 references each outermost 20-MHz carrier against a 20-MHz window offset 22\,MHz outward. EVM is computed over all populated subcarriers of one capture from QAM symbols recovered after time/common-phase alignment and OFDM demodulation, without downsampling or averaging.

\subsection{Neural Network Training}
\label{sec:training}
Training follows the two-step direct-learning flow of \texttt{OpenDPDv2}~\cite{Wu2025OpenDPDv2,OpenDPDSoftware}, while IPG-DWTCN and its hardware datapath are developed here. For each DUT, a GRU PA surrogate is trained via time-domain mean-squared error (MSE), selected by validation normalized MSE (NMSE), and frozen across all DPD runs. Each DPD is cascaded ahead of this surrogate and trained end-to-end against $Gx[n]$, where $G$ is the training-split peak output-to-input amplitude ratio.

All runs use Table~\ref{tab:experimental_setup} with evaluation batch size 256, global-norm gradient clipping at 200, and AdamW weight decay 0.01. The learning rate (LR) halves after 10 epochs without validation gain down to $10^{-6}$, monitoring NMSE for PA modeling and mean ACPR for DPD. Each DPD trains for 100 epochs over seeds 0-4, keeping the best-validation-ACPR checkpoint per seed.

Near a 200-parameter budget, topologies and hyperparameters remain fixed across DUT1-DUT3: noncausal TCNN~\cite{Duan2025TCN-DPD} ($C=7$), dense GRU (DGRU, $H=4$), BO-JANET ($H=1$, 16-tap window, 6 vector-demodulator branches), augmented phase-normalized residual recurrent unit (APNRRU, $H=1$, 16-tap window, 3 finite impulse response (FIR) branches, widths 7/3), and DVR-JANET ($H=5$, 3 DVR units), alongside L3/C9 ($K=5$) IPG-DWTCN. BO-JANET on DUT3 is the sole exception: since scalar-state $H=1$ (224 parameters) lacks memory capacity for the 3.50-GHz GaN Doherty PA under 200-MHz ten-carrier drive and fails to beat no-DPD, $H=2$ (258 parameters) is used (Table~\ref{tab:model-comparison}).

Complexity is counted as static forward-graph FLOPs per complex output sample (2 per MAC, 1 per bias or element-wise arithmetic operation; shifts, comparisons, saturation, and data movement excluded). IPG-DWTCN ($K=5$) uses $5+22C+13LC$ FLOPs/sample: 5 I/Q operations ($|x[n]|^2$ and output skip), $22C$ front/back-end operations ($7C$ input, $2C$ \texttt{Hardswish8}, $3C$ gate, $5C$ two-branch \texttt{Quadsigmoid}, $C$ gating product, $4C$ output), and $13LC$ temporal operations ($10$ DW MAC, $2$ \texttt{Hardswish8}, $1$ residual add per PE), giving 554 FLOPs/sample at L3/C9.

\subsection{Hardware Implementation and Evaluation Setup}
\label{sec:hw-setup}
The L3/C9 mixed-precision architecture (Table~\ref{tab:fixed-formats}) is described in SystemVerilog and realized as a four-lane ($P=4$) FPGA core for real-time HIL testing and a single-lane ($P=1$) 12-nm ASIC core. Cadence Xcelium 25.03 verifies the RTL bit-accurately against the quantized PyTorch model over all 35,333 DUT1 validation samples.

For FPGA HIL, Vivado 2025.2 implements the seed-0 $P=4$ core on the XCZU67DR RFSoC at 320\,MHz (128-bit W12A16 bus, 1.28\,GSa/s). Direct memory access (DMA) replay unpacks each 256-bit, eight-sample AXI-Stream beat into two core beats; the reverse packer feeds the RF digital-to-analog converter (RF-DAC), while the RF analog-to-digital converter (RF-ADC) returns captures to host memory via receive DMA. Post-route resource reports cover the arithmetic core, excluding board-interface and parameter-loading logic. Real-time signal quality is referenced to the $-38$-dB ($1.26\%$) IEEE~802.11be MCS13 EVM limit~\cite{Breinbauer2024EHTEVM}.

For ASIC evaluation, Cadence Genus 23.15 synthesizes the $P=1$ core in GlobalFoundries 12LP+, Innovus 23.35 performs placement, clock-tree synthesis, routing, multi-corner multi-mode timing, and power analysis, and integrated Quantus 23.1.1 extracts post-layout parasitics. For power analysis, Xcelium captures 4096 steady-state DUT1 cycles to a waveform database, Genus maps the activity to switching activity interchange format (SAIF), and Innovus propagates it through the routed design at slow-slow (SSG, 0.72\,V, $125\,^{\circ}$C), TT (0.80\,V, $25\,^{\circ}$C), and fast-fast (FFG, 0.88\,V, $-40\,^{\circ}$C) corners. Routed-netlist Xcelium simulations additionally generate value change dump (VCD) activity and verify all 35,333 samples with maximum-delay standard delay format (SDF) annotation.

%======================================================================
\section{Experimental Results and Discussion}
\label{sec:results}
%======================================================================

We first ablate gating, activations, and channel width on DUT1 and benchmark the frozen L3/C9 model offline across all three DUTs, then report four-lane FPGA HIL and 12-nm ASIC results before comparing against the state of the art.

\subsection{Ablation Study of IPG-DWTCN}
\label{sec:ablation}
\begin{table}[!t]
\centering
\caption{Measured gating comparison at three near-matched parameter budgets for $L=3$; $C$ is the actual feature-channel count of each row.}
\label{tab:gating-ablation}
\scriptsize
\setlength{\tabcolsep}{0.7pt}
\renewcommand{\arraystretch}{1.06}
\begin{threeparttable}
\begin{tabularx}{\columnwidth}{|>{\centering\arraybackslash}p{0.09\columnwidth}|>{\centering\arraybackslash}p{0.14\columnwidth}|>{\centering\arraybackslash}p{0.12\columnwidth}|AE|}
\hline
$\boldsymbol{C}$ & \textbf{Gate} & \textbf{\#params} &
\begin{tabular}[c]{@{}c@{}}\textbf{ACPR}\\\textbf{(dBc)}\end{tabular} &
\begin{tabular}[c]{@{}c@{}}\textbf{EVM}\\\textbf{(dB)}\end{tabular} \\ \hline\hline
8 & No gate & 168 & \MeanStd{-43.39}{0.52} & \MeanStd{-39.27}{0.42} \\
7 & Amplitude & 161 & \MeanStd{-45.12}{0.65} & \MeanStdPct{-41.20}{1.01}{0.87} \\
7 & Power & 161 & \MeanStd{-45.11}{0.52} & \MeanStdPct{-41.13}{0.91}{0.88} \\ \hline
9 & No gate & 189 & \MeanStd{-43.23}{0.67} & \MeanStd{-40.09}{0.43} \\
8 & Amplitude & 184 & \MeanStd{-45.60}{0.05} & \MeanStdPct{-41.86}{0.24}{0.81} \\
8 & Power & 184 & \MeanStd{-45.52}{0.28} & \MeanStdPct{-41.70}{0.78}{0.82} \\ \hline
10 & No gate & 210 & \MeanStd{-44.50}{1.16} & \MeanStd{-41.52}{0.98} \\
9 & Amplitude & 207 & \MeanStd{-45.66}{0.23} & \MeanStdPct{-41.91}{0.44}{0.80} \\
\textbf{9} & \textbf{Power} & \textbf{207} & \BoldMeanStd{-45.90}{0.05} & \BoldMeanStdPct{-42.91}{0.11}{0.72} \\ \hline
\end{tabularx}
\end{threeparttable}
\end{table}

Tables~\ref{tab:gating-ablation} and~\ref{tab:activation-ablation} evaluate feature gating and activation choices on DUT1 at $L=3$. To separate gating benefits from parameter growth, Table~\ref{tab:gating-ablation} pairs gated models at $C\in\{7,8,9\}$ (161, 184, and 207 parameters) with wider un-gated DW-TCNs at $C\in\{8,9,10\}$ (168, 189, and 210 parameters). Across the three budgets, instantaneous-power gating improves mean ACPR by 1.72, 2.29, and 1.40\,dB and mean EVM by 1.86, 1.61, and 1.39\,dB while using 7, 5, and 3 fewer parameters, confirming that envelope-driven channel modulation is far more effective than simply widening the convolution stack.

Comparing gating variables in Table~\ref{tab:gating-ablation}, $|x[n]|^2$-driven (power) and $|x[n]|$-driven (amplitude) gates perform comparably at $C=7$ and $C=8$, whereas at $C=9$ the power gate improves mean ACPR by 0.24\,dB and mean EVM by 1.00\,dB (reaching $-45.90\pm0.05$\,dBc and $-42.91\pm0.11$\,dB) with tighter seed consistency. Because $|x[n]|^2$ is already available in $\mathbf{f}[n]$ without a square-root or CORDIC unit, instantaneous power offers both superior RF fidelity and simpler hardware.

\begin{table}[!t]
\centering
\caption{Measured gate- and main-path-activation combinations for the $L=3$, $C=9$, 207-parameter model across five random seeds.}
\label{tab:activation-ablation}
\TblFont
\setcellgapes{1pt}\makegapedcells
\setlength{\tabcolsep}{1.5pt}
\renewcommand{\arraystretch}{\TblStretch}
\begin{tabularx}{\columnwidth}{|>{\centering\arraybackslash}p{0.24\columnwidth}|>{\centering\arraybackslash}p{0.175\columnwidth}|AE|}
\hline
\textbf{Gate activation} & \begin{tabular}[c]{@{}c@{}}\textbf{Main-path}\\\textbf{activation}\end{tabular} &
\begin{tabular}[c]{@{}c@{}}\textbf{ACPR}\\\textbf{(dBc)}\end{tabular} &
\begin{tabular}[c]{@{}c@{}}\textbf{EVM}\\\textbf{(dB)}\end{tabular} \\ \hline\hline
\MR{2}{Logistic sigmoid} & HardSwish & \MeanStd{-45.76}{0.07} & \MeanStdPct{-42.49}{0.49}{0.75} \\
 & Hardswish8 & \MeanStd{-45.87}{0.13} & \MeanStdPct{-42.61}{0.57}{0.74} \\ \hline
\MR{2}{Hard sigmoid} & HardSwish & \MeanStd{-45.54}{0.22} & \MeanStdPct{-41.96}{0.52}{0.80} \\
 & Hardswish8 & \MeanStd{-45.74}{0.17} & \MeanStdPct{-42.13}{0.56}{0.78} \\ \hline
\MR{2}{\makecell{Smoothed hard\\sigmoid}} & HardSwish & \MeanStd{-45.74}{0.14} & \MeanStdPct{-42.35}{0.19}{0.76} \\
 & Hardswish8 & \MeanStd{-45.75}{0.18} & \MeanStdPct{-42.12}{0.45}{0.78} \\ \hline
Quadsigmoid & HardSwish & \MeanStd{-45.60}{0.28} & \MeanStdPct{-42.35}{0.90}{0.76} \\
\textbf{Quadsigmoid} & \textbf{Hardswish8} & \BoldMeanStd{-45.90}{0.05} & \BoldMeanStdPct{-42.91}{0.11}{0.72} \\ \hline
\end{tabularx}
\end{table}

In the L3/C9 activation study (Table~\ref{tab:activation-ablation}), piecewise-linear gates (hard sigmoid and smoothed hard sigmoid) degrade mean EVM by $0.56$-$0.95$\,dB relative to \texttt{Quadsigmoid}/\texttt{Hardswish8} due to slope discontinuities. Conversely, pairing $C^1$-continuous \texttt{Quadsigmoid} with shift-scaled \texttt{Hardswish8} yields the best mean ACPR ($-45.90$\,dBc) and EVM ($-42.91$\,dB) among all eight pairs, matching logistic sigmoid/\texttt{Hardswish8} within 0.03\,dB ACPR while improving EVM by 0.30\,dB and eliminating LUTs and $\div6$ dividers.

\subsection{Model Size Sweep}
\label{sec:size-sweep}

\begin{figure}[!t]
\centering
\includegraphics[width=\columnwidth]{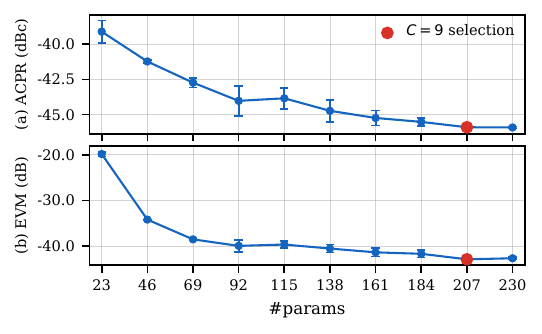}
\caption{Measured performance-parameter trade-off of the $L=3$ IPG-DWTCN width sweep ($C\in\{1,\dots,10\}$): (a)~ACPR and (b)~EVM. Markers and error bars denote five-seed means and standard deviations; the $C=9$ (207-parameter) point reports held-out test data as in Tables~\ref{tab:gating-ablation}-\ref{tab:activation-ablation}.}
\label{fig:width-tradeoff}
\end{figure}
Fig.~\ref{fig:width-tradeoff}(a) and~(b) sweep channel width $C\in\{1,\dots,10\}$ (23 to 230 parameters at 23 parameters/channel) on DUT1. Linearization improves rapidly from $C=1$ to $C=4$ (23 to 92 parameters), reducing ACPR from $-39.11$ to $-44.02$\,dBc and EVM from $-19.82$ to $-39.96$\,dB, before tapering at larger widths. Guided by the 45-dB adjacent-channel leakage power ratio (ACLR) requirement for NR base stations in 3GPP Technical Specification (TS)~38.104 Section~6.6.3~\cite{3gpp38104}, we adopt $-45$\,dBc ACPR as a design target under the protocol of Section~\ref{sec:setup}. Although $C=7$ and $C=8$ cross $-45$\,dBc in mean ACPR, their seed spread remains wider; $C=9$ (207 parameters) is selected on validation data as the smallest width consistently clearing $-45$\,dBc across all five seeds, confirmed on held-out test captures ($-45.90\pm0.05$\,dBc ACPR, $-42.91\pm0.11$\,dB EVM) with negligible gain at $C=10$ (230 parameters).

\subsection{Offline RF Comparison with Prior Models}
\label{sec:comparison}
Table~\ref{tab:model-comparison} benchmarks the frozen L3/C9 IPG-DWTCN against five neural baselines near 200 parameters across DUT1-DUT3 under offline RF playback (real-time on-board HIL results follow in Section~\ref{sec:hwresults}). Compared with our prior noncausal 203-parameter TCNN~\cite{Duan2025TCN-DPD} (which uses look-ahead without envelope gating), causal IPG-DWTCN improves mean ACPR/EVM by 1.11/2.03, 1.34/0.73, and 4.01/1.54\,dB on DUT1-DUT3, respectively, while lowering complexity from 620 to 554 FLOPs/sample.

\begin{table*}[!t]
\centering
\caption{Physical-DUT RF measurements using offline-generated waveforms for DPD models near 200 parameters; results are mean $\pm$ standard deviation across five random seeds.}
\label{tab:model-comparison}
\TblFont
\setcellgapes{1pt}\makegapedcells
\setlength{\tabcolsep}{1pt}
\renewcommand{\arraystretch}{\TblStretch}
\begin{threeparttable}
\begin{tabularx}{\textwidth}{|>{\centering\arraybackslash}p{0.145\textwidth}|>{\centering\arraybackslash}p{0.062\textwidth}|>{\centering\arraybackslash}p{0.058\textwidth}|*{6}{>{\centering\arraybackslash}X}|}
\hline
\multirow{2}{*}{\textbf{Model}} & \multirow{2}{*}{\textbf{\#params}} & \multirow{2}{*}{\makecell{\textbf{Alg.}\\\textbf{FLOPs}\tnote{a}}} &
\multicolumn{2}{c|}{\textbf{DUT1 (6.11\,GHz)}} & \multicolumn{2}{c|}{\textbf{DUT2 (2.40\,GHz)}} & \multicolumn{2}{c|}{\textbf{DUT3 (3.50\,GHz)}} \\ \cline{4-9}
 & & & \textbf{ACPR (dBc)} & \textbf{EVM (dB)} & \textbf{ACPR (dBc)} & \textbf{EVM (dB)} & \textbf{ACPR (dBc)} & \textbf{EVM (dB)} \\ \hline\hline
w/o DPD & - & - & $-35.40$ & $-33.61$ & $-35.97$ & $-30.93$ & $-21.92$ & $-19.97$ \\ \hline
TCN~\cite{Duan2025TCN-DPD} & 203 & 620 & \MeanStd{-44.79}{0.88} & \MeanStdPct{-40.88}{1.38}{0.90} & \MeanStd{-45.50}{0.41} & \MeanStdPct{-43.21}{0.63}{0.69} & \MeanStd{-41.38}{0.81} & \MeanStdPct{-38.31}{1.20}{1.22} \\
DGRU~\cite{wu2024opendpd} & 186 & 431 & \MeanStd{-42.81}{0.67} & \MeanStdPct{-38.44}{0.32}{1.20} & \MeanStd{-45.97}{0.47} & \MeanStdPct{-43.22}{0.38}{0.69} & \MeanStd{-42.57}{0.41} & \MeanStdPct{-38.50}{0.57}{1.19} \\
BO-JANET~\cite{BOJANET}\tnote{b} & 224 & 924 & \MeanStd{-41.29}{0.46} & \MeanStdPct{-36.50}{0.19}{1.50} & \MeanStd{-42.94}{0.25} & \MeanStdPct{-39.74}{0.20}{1.03} & \MeanStd{-40.64}{0.84} & \MeanStdPct{-35.69}{0.99}{1.64} \\
APNRRU~\cite{APNRRU} & 242 & 792 & \MeanStd{-45.77}{0.34} & \MeanStdPct{-42.53}{1.01}{0.75} & \MeanStd{-46.29}{0.34} & \MeanStdPct{-43.94}{0.29}{0.64} & \MeanStd{-45.05}{0.51} & \MeanStdPct{-38.86}{1.31}{1.14} \\
DVR-JANET~\cite{DVRJANET} & 215 & 647 & \MeanStd{-41.92}{1.40} & \MeanStdPct{-38.08}{1.13}{1.25} & \MeanStd{-45.21}{0.38} & \MeanStdPct{-43.25}{0.92}{0.69} & \MeanStd{-42.11}{0.92} & \MeanStdPct{-36.97}{0.62}{1.42} \\
\textcolor[HTML]{CB0000}{\textbf{IPG-DWTCN} (ours)} & \textcolor[HTML]{CB0000}{\textbf{207}} & \textcolor[HTML]{CB0000}{\textbf{554}} &
\textcolor[HTML]{CB0000}{\BoldMeanStd{-45.90}{0.05}} & \textcolor[HTML]{CB0000}{\BoldMeanStdPct{-42.91}{0.11}{0.72}} &
\textcolor[HTML]{CB0000}{\BoldMeanStd{-46.84}{0.06}} & \textcolor[HTML]{CB0000}{\BoldMeanStdPct{-43.94}{0.09}{0.64}} &
\textcolor[HTML]{CB0000}{\BoldMeanStd{-45.39}{0.54}} & \textcolor[HTML]{CB0000}{\BoldMeanStdPct{-39.85}{0.62}{1.02}} \\ \hline
\end{tabularx}
\begin{tablenotes}
\item[a] Algorithmic FLOPs per output complex I/Q sample, counted as defined in Section~\ref{sec:training}.
\item[b] BO-JANET uses $H=1$ (224 parameters, 924 FLOPs/sample) on DUT1/2 and $H=2$ (258 parameters, 1010 FLOPs/sample) on DUT3 because $H=1$ fails to improve on the undistorted baseline there.
\end{tablenotes}
\end{threeparttable}
\end{table*}

\begin{figure*}[!t]
\centering
\includegraphics[width=0.98\textwidth]{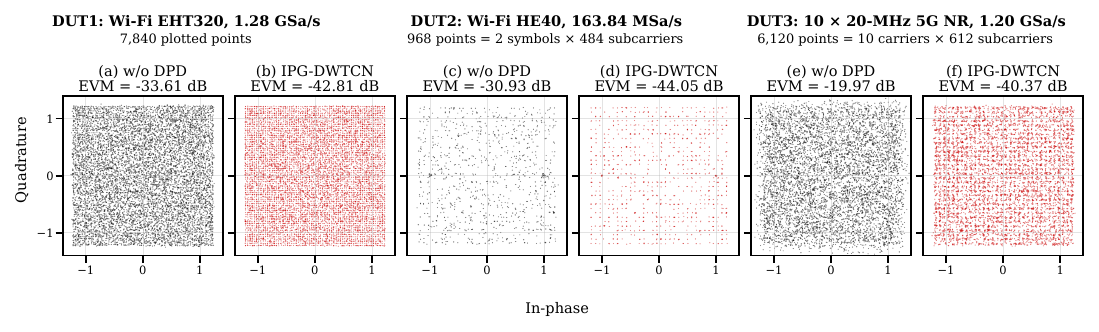}
\caption{Measured constellations without DPD and with IPG-DWTCN for DUT1 (EHT-320), DUT2 (HE-40), and DUT3 (ten-carrier 5G~NR). Each panel reports the EVM of the plotted capture.}
\label{fig:constellation}
\end{figure*}

On the 320-MHz 4096-QAM Wi-Fi~7 testbed (DUT1), bit-accurate mixed-precision IPG-DWTCN improves ACPR from $-35.40$ to $-45.90\pm0.05$\,dBc and EVM from $-33.61$ to $-42.91\pm0.11$\,dB, outperforming all five FP32 baselines in both mean accuracy and seed stability ($7\times$-$28\times$ smaller ACPR standard deviation). Recurrent APNRRU~\cite{APNRRU} comes closest ($-45.77$\,dBc ACPR, $-42.53$\,dB EVM), but requires 35 more parameters (242 versus 207), 238 more FLOPs/sample (554 is 30.1\% below 792), and state feedback precluding II~$=1$ streaming.

With a 40-MHz signal on the Wi-Fi~6 front-end module (DUT2), leading models cluster more closely, yet IPG-DWTCN still achieves the best mean ACPR ($-46.84\pm0.06$\,dBc, beating APNRRU by 0.55\,dB) and ties APNRRU for the best mean EVM ($-43.94$\,dB), delivering 10.87-dB ACPR and 13.01-dB EVM gains relative to operation without DPD.

DUT3 (35.20-dBm GaN Doherty PA, ten-carrier 200-MHz 5G~NR) exhibits the severest raw distortion ($-21.92$\,dBc ACPR, $-19.97$\,dB EVM) from main-to-peaking amplifier transition and trapping memory. Without retuning the DUT1-selected architecture, IPG-DWTCN reaches $-45.39\pm0.54$\,dBc ACPR and $-39.85\pm0.62$\,dB EVM, providing 23.47- and 19.88-dB gains and surpassing runner-up APNRRU ($-45.05$\,dBc, $-38.86$\,dB) by 0.34\,dB in ACPR and 0.99\,dB in EVM. IPG-DWTCN thus achieves the best mean ACPR and best or tied mean EVM across all three RF devices.

\begin{figure*}[!t]
\begin{minipage}[t]{0.48\textwidth}
\centering
\includegraphics[width=\linewidth]{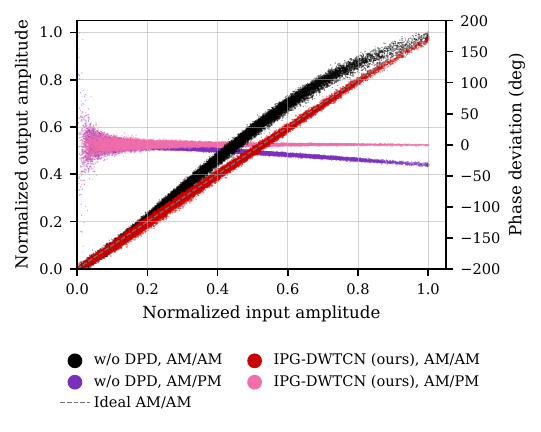}
\caption{Measured AM/AM and AM/PM characteristics of DUT3 with and without IPG-DWTCN DPD. Complex gain is referenced to the lowest-amplitude 5\% of samples.}
\label{fig:amam-ampm-dut3}
\end{minipage}
\hfill
\begin{minipage}[t]{0.48\textwidth}
\centering
\includegraphics[width=\linewidth]{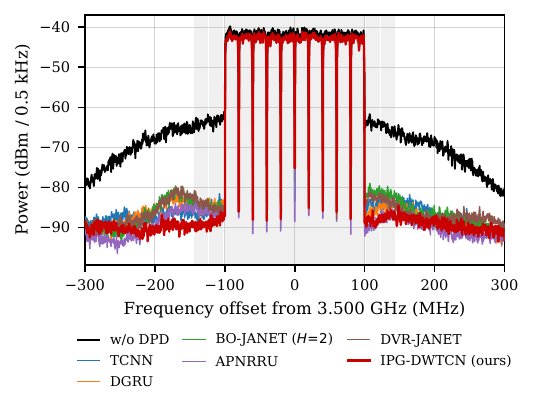}
\caption{Measured output PSD of DUT3 without DPD and with each model in Table~\ref{tab:model-comparison}; shaded bands mark the edge-carrier reference and adjacent ACPR windows.}
\label{fig:psd-performance-dut3}
\end{minipage}
\end{figure*}

Fig.~\ref{fig:constellation} compares single-capture constellations without DPD and with IPG-DWTCN on DUT1 ((a)-(b)), DUT2 ((c)-(d)), and DUT3 ((e)-(f)), showing tightly restored QAM clusters consistent with the five-seed EVM means in Table~\ref{tab:model-comparison}. On DUT3, Fig.~\ref{fig:amam-ampm-dut3} confirms that IPG-DWTCN linearizes amplitude-to-amplitude (AM/AM) compression and flattens the nearly $25^\circ$ amplitude-to-phase (AM/PM) lag, while the output power spectral densities (PSDs) in Fig.~\ref{fig:psd-performance-dut3} demonstrate $>23$\,dB adjacent-channel suppression across both sidebands.

\subsection{Hardware Implementation Results}
\label{sec:hwresults}

\subsubsection{FPGA HIL Validation}
\label{sec:fpga-results}
Under the four-lane polyphase setup of Section~\ref{sec:hw-setup}, the post-route XCZU67DR core meets setup and hold timing at the 344.83-MHz (2.90-ns) constraint and runs at 320\,MHz (3.13-ns period) for continuous 1.28-GSa/s streaming. In real-time on-board HIL operation on DUT1 with the 320-MHz EHT signal, the W12A16-interface L3/C9 core achieves $-45.82$\,dBc ACPR and $-42.94$\,dB ($0.71\%$) EVM (Table~\ref{tab:soa}), matching the offline five-seed benchmark in Table~\ref{tab:model-comparison} within 0.08\,dB ACPR and 0.03\,dB EVM with a 4.94-dB margin against the MCS13 EVM limit~\cite{Breinbauer2024EHTEVM}. Table~\ref{tab:fpga} shows that post-route utilization peaks at 52.35\% of DSP48E2 slices (980 DSPs, 11.30\% LUTs, and 10.16\% FFs) with 0\% block RAM (BRAM) or UltraRAM (URAM), confirming strictly RAM-free polyphase execution.

% Cappello validation evidence: author-posted abstract, accessed 2026-11-13.
% https://independent.academia.edu/GautamJindal2
\begin{table*}[!t]
\centering
\caption{Comparison with reported DPD hardware and signal quality, including the validation path for the reported RF metrics. This work combines 12-nm FinFET ASIC post-layout estimates with on-board FPGA HIL measurements.}
\label{tab:soa}
\TblFont
\renewcommand{\arraystretch}{1.18}
\setlength{\tabcolsep}{1pt}
\nomakegapedcells
\begin{threeparttable}
\fontsize{7.5}{9}\selectfont
% Distribute spare width as cell padding so partial rules meet the borders.
\newcommand{\SoATabular}{%
\begin{tabular}{|c|cccccccc|ccc|l|}
\hline
\multirow{2}{*}[-4.5pt]{\textbf{Work}} & \multicolumn{8}{c|}{\textbf{Model and hardware}} & \multicolumn{4}{c|}{\textbf{Signal quality and validation}} \\ \cline{2-13}
 & \makecell{\textbf{Platform}\\\textbf{(nm)}} & \textbf{Model} & \makecell{\textbf{Precision}\\\textbf{(bits)}} & \textbf{\#params} & \makecell{$\boldsymbol{f_s}$\\\textbf{(MSa/s)}} & \makecell{\textbf{Latency}\\\textbf{(ns)}} & \makecell{\textbf{Power}\\\textbf{(W)}} & \makecell{\textbf{Energy/sample}\\\textbf{(pJ)}}\tnote{1} & \makecell{$\boldsymbol{f_{\mathrm{BB}}}$\\\textbf{(MHz)}} & \makecell{\textbf{ACPR}\\\textbf{(dBc)}} & \makecell{\textbf{EVM}\\\textbf{(dB)}} & \makecell{\textbf{RF validation}\\\textbf{path}} \\ \hline\hline
\textbf{This work} & ASIC, 12 & IPG-DWTCN & W12A16\tnote{2} & 207 & 1280 & 40.63 & 0.14 & 111.10 & 320 & $-45.82$ & $-42.94$ & HIL (FPGA) \\
\cite{Li2025ISCAS} & ASIC, 22 & GRU & W12A12 & 502 & 250 & 7.50 & 0.20 & 780.00 & 60 & $-45.30$ & $-39.80$ & RF playback \\
\cite{Huang2020} & FPGA, 16 & ECRV & W?A16 & 36 & 1200 & -- & -- & -- & 400 & $-44.70$ & $-39.20$ & Conducted HIL \\
\cite{Huang2020}\tnote{3} & FPGA, 16 & ECRV & W?A16 & 36 & 2400 & -- & 0.958 & 399 & 800 & $-40.55$ & $-38.42$ & OTA HIL \\
\cite{Cappello2022} & FPGA, 28 & MP & W?A16 & 9 & 250 & 40 & 0.24 & 960 & 20 & $-49.00$ & -- & HIL unconfirmed \\
\cite{Yue2022} & FPGA, 28 & GMP & W?A16 & 38 & 400 & -- & 0.89 & 2225 & 100 & $-46.45$ & -- & RF playback \\
\cite{10994208}\tnote{4} & GPU, 5 & PN-TDNN & FP32 & 909 & $\sim$1000 & -- & -- & -- & 200 & $-45.10$ & $-35.55$ & RF + timing \\
\cite{11449061} & FPGA, -- & Sparse PNTDNN & W14A14 & 64 & 170 & -- & 0.405 & 2382 & 20 & $-59.40$ & $-54.00$ & Simulation \\
\hline
\end{tabular}%
}
\sbox0{\SoATabular}
\addtolength{\tabcolsep}{\dimexpr(\textwidth-\wd0)/26\relax}
\SoATabular
\begin{tablenotes}
\item[1] Energy/sample is power divided by $f_s$, using unrounded power when available. ``?'' indicates partially reported precision and ``--'' an unreported value. ACPR/EVM depend on the PA, waveform, and validation method.
\item[2] Internal arithmetic uses operator-specific widths as detailed in Section~\ref{sec:operators}.
\item[3] Eight-lane envelope-based complex reduced Volterra (ECRV) configuration. Power is the estimated dynamic component. OTA denotes over-the-air testing.
\item[4] Throughput is from CUDA batch timing. RF metrics are from the 40.6-dBm PA output operating point.
\end{tablenotes}
\end{threeparttable}
\end{table*}

\begin{table}[!t]
\centering
\caption{Post-route results of the four-lane, mixed-precision FPGA compute core on XCZU67DR.}
\label{tab:fpga}
\TblFont
\setcellgapes{1pt}\makegapedcells
\setlength{\tabcolsep}{\TblColSep}
\renewcommand{\arraystretch}{\TblStretch}
\begin{tabularx}{\columnwidth}{|>{\centering\arraybackslash}X|>{\centering\arraybackslash}X|>{\centering\arraybackslash}X|>{\centering\arraybackslash}X|}
\hline
\textbf{Resource} & \textbf{Used} & \textbf{Available} & \textbf{Utilization} \\
\hline\hline
LUT & 25\,270 & 223\,680 & 11.30\% \\
FF & 45\,459 & 447\,360 & 10.16\% \\
DSP48E2 & 980 & 1\,872 & 52.35\% \\
BRAM & 0 & 648 & 0.00\% \\
URAM & 0 & 160 & 0.00\% \\
\hline
\end{tabularx}
\end{table}

\subsubsection{12-nm ASIC Post-Layout Evaluation}
\label{sec:asic-results}
\begin{figure}[!t]
    \centering
    \subfloat[]{\includegraphics[width=0.36\linewidth]{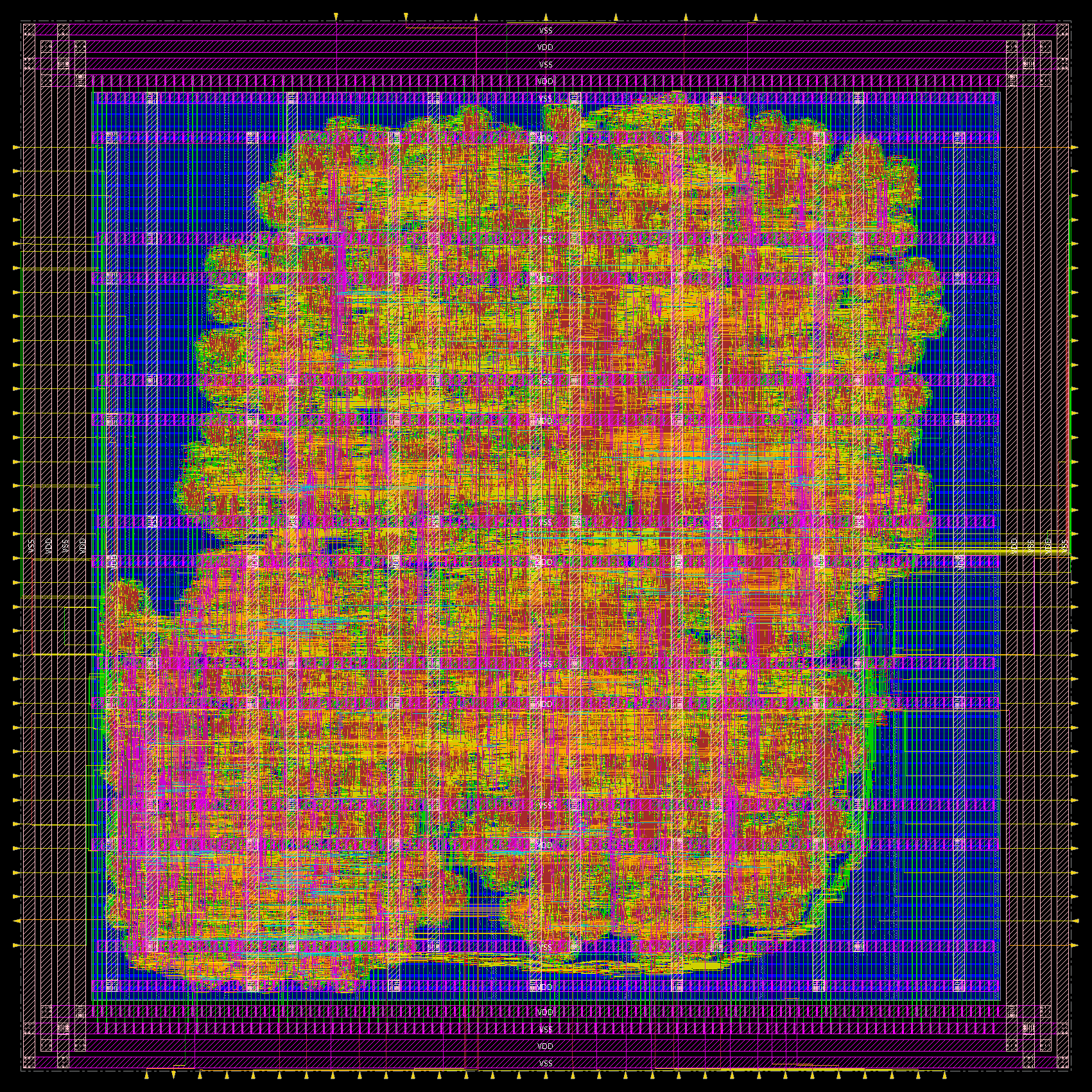}\label{fig:layout-die}}\hfill
    \subfloat[]{\includegraphics[width=0.61\linewidth]{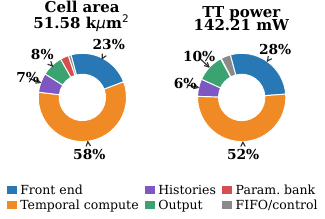}\label{fig:layout-cost}}
    \caption{Single-lane L3/C9 accelerator: (a) routed 12-nm FinFET layout; (b) functional-cell area and TT power breakdowns. The core measures $321.30\times321.12\,\mu$m$^2$ within a $371.36\times371.20\,\mu$m$^2$ die.}
    \label{fig:layout}
\end{figure}

Fig.~\ref{fig:layout}(a) shows the routed 12-nm FinFET layout of the single-lane ($P=1$) programmable L3/C9 core, occupying a 0.10-mm$^2$ core (0.14-mm$^2$ die), and Table~\ref{tab:asic} summarizes post-layout signoff metrics. Meeting setup and hold timing across all evaluated corners under the 0.78-ns (1.28-GHz) constraint, the core delivers 1.28\,GSa/s at II~$=1$ with a deterministic 52-cycle (40.63-ns) latency.

Using the routed-netlist switching-activity flow of Section~\ref{sec:hw-setup}, TT power at 1.28\,GHz is 142.21\,mW (111.10\,pJ per complex sample); SSG power drops to 127.90\,mW (99.92\,pJ/sample) due to the lower 0.72-V supply, whereas FFG power reaches 166.41\,mW (130.01\,pJ/sample). In Fig.~\ref{fig:layout}(b), the 27 temporal PEs and FF causal histories account for 58\% and 7\% of cell area (totaling 64.73\% of 51.58\,k$\mu$m$^2$) and 52\% and 6\% of TT power (totaling 57.99\%), followed by the gated front end (23\% area, 28\% power) and output back end (8\% area, 10\% power), with under 5\% in the parameter bank and FIFO/control. Channel-separated DW convolutions avoid dense $C\times C$ temporal matrices, and progressive 10/9/8-bit weights cut DW1-DW3 MAC cost.

\subsection{Comparison with the State of the Art}
\label{sec:soa}
Table~\ref{tab:soa} compares hardware and RF performance, pairing our 12-nm post-layout ASIC estimates with FPGA HIL measurements of the same fixed-point datapath.

\emph{Model and hardware efficiency:} IPG-DWTCN uses 207 parameters, compared with 502 for the prior 22-nm GRU ASIC~\cite{Li2025ISCAS} and 909 for the pruned phase-normalized TDNN on GPU~\cite{10994208}. Removing recurrent feedback increases throughput from 250\,MSa/s to 1.28\,GSa/s ($5.1\times$) and reduces energy from 780.00 to 111.10\,pJ/sample ($7.0\times$). Reported FPGA energies span 399--2382\,pJ/sample ($3.6\times$--$21.4\times$ our ASIC estimate), from eight-lane ECRV dynamic power~\cite{Huang2020} to SparseDPD total power~\cite{11449061}.

\emph{Signal quality and validation:} On DUT1, 1.28-GSa/s FPGA HIL with 320-MHz 4096-QAM achieves $-45.82$\,dBc ACPR and $-42.94$\,dB EVM, at $5.3\times$ the reported bandwidth of our 60-MHz GRU ASIC~\cite{Li2025ISCAS}. Huang et al.~\cite{Huang2020} report $-44.70$\,dBc ACPR and $-39.20$\,dB EVM for their conducted 400-MHz test at 1.2\,GSa/s. SparseDPD~\cite{11449061} evaluates RF performance in simulation with a 20-MHz signal. LW-ResGRU~\cite{11278888} evaluates offline datasets and FPGA high-level synthesis (HLS).

% Keep the ASIC summary after the comparison and FPGA tables in reading order.
\begin{table}[!t]
\centering
\caption{Post-layout results of the single-lane 12-nm FinFET accelerator.}
\label{tab:asic}
\TblFont
\nomakegapedcells
\setlength{\extrarowheight}{1pt}
\setlength{\tabcolsep}{\TblColSep}
\renewcommand{\arraystretch}{1.12}
\renewcommand{\tabularxcolumn}[1]{m{#1}}
\begin{tabularx}{\columnwidth}{|>{\centering\arraybackslash}X|>{\centering\arraybackslash}m{0.64\columnwidth}|}
\hline
\textbf{Metric} & \textbf{Result} \\
\hline\hline
Technology & 12-nm FinFET \\ \hline
Clock frequency & 1.28\,GHz (0.78\,ns constraint) \\ \hline
Throughput & 1.28\,GSa/s, II~$=1$ \\ \hline
Latency  & 52 cycles (40.63\,ns) \\ \hline
Core / die area & 0.10 / 0.14\,mm$^2$ \\ 
\hline\hline
\multicolumn{2}{|c|}{\textbf{Activity-based routed power}} \\ \hline\hline
SSG, 0.72\,V, $125\,^{\circ}$C & 127.90\,mW, 99.92\,pJ/sample \\ 
TT, 0.80\,V, $25\,^{\circ}$C & 142.21\,mW, 111.10\,pJ/sample \\ 
FFG, 0.88\,V, $-40\,^{\circ}$C & 166.41\,mW, 130.01\,pJ/sample \\
\hline
\end{tabularx}

\end{table}

%======================================================================
\section{Conclusion}
\label{sec:conclusion}
%======================================================================
This paper presented \texttt{StreamingDPD}, an algorithm-hardware co-designed streaming accelerator for wideband GSa/s DPD. Replacing recurrence with $|x[n]|^2$-driven channel gating, channel-separated causal dilated DW convolutions, shift-friendly quadratic activations, and layer-progressive mixed precision, the 207-parameter IPG-DWTCN cuts computation to 554 FLOPs/sample while achieving the best mean ACPR ($-46.84$ to $-45.39$\,dBc) and best or tied mean EVM ($-43.94$ to $-39.85$\,dB) across three wideband RF devices. Four-lane RFSoC HIL on 320-MHz 4096-QAM Wi-Fi~7 reaches $-45.82$\,dBc ACPR and $-42.94$\,dB EVM at 1.28\,GSa/s with zero BRAM/URAM, while single-lane 12-nm FinFET post-layout evaluation yields a 0.10-mm$^2$ core, 52-cycle (40.63-ns) latency at II~$=1$, and 142.21-mW TT power (111.10\,pJ/sample), improving throughput by $5.1\times$ and energy efficiency by $7.0\times$ over the prior neural DPD ASIC. Future work will pursue silicon fabrication, shadow-bank background adaptation, and multi-antenna MIMO transmitter integration.

\section*{Acknowledgment}
The authors thank OnMicro for providing the 2.40-GHz \mbox{Wi-Fi~6} front-end module (DUT2) and Ampleon for providing the 3.50-GHz GaN Doherty PA (DUT3) used in this work. We also thank GlobalFoundries
for providing us the 12LP-PLUS PDK through the GF12+
University Partnership Programme.
% \balance
\bibliographystyle{IEEEtran}
\bibliography{refs}

@article{GMP,
  author    = {D. R. Morgan and Z. Ma and J. Kim and M. G. Zierdt and J. Pastalan},
  title     = {A Generalized Memory Polynomial Model for Digital Predistortion of {RF} Power Amplifiers},
  journal   = {IEEE Trans. Signal Process.},
  volume    = {54},
  number    = {10},
  pages     = {3852--3860},
  month     = oct,
  year      = {2006}
}

@article{VDLSTM,
  author    = {H. Li and Y. Zhang and G. Li and F. Liu},
  title     = {Vector Decomposed Long Short-Term Memory Model for Behavioral Modeling and Digital Predistortion for Wideband {RF} Power Amplifiers},
  journal   = {IEEE Access},
  volume    = {8},
  pages     = {63780--63789},
  year      = {2020}
}

@article{PGJANET,
  author    = {T. Kobal and Y. Li and X. Wang and A. Zhu},
  title     = {Digital Predistortion of {RF} Power Amplifiers With Phase-Gated Recurrent Neural Networks},
  journal   = {IEEE Trans. Microw. Theory Techn.},
  volume    = {70},
  number    = {6},
  pages     = {3291--3299},
  month     = jun,
  year      = {2022}
}

@article{DVRJANET,
  author    = {T. Kobal and A. Zhu},
  title     = {Digital Predistortion of {RF} Power Amplifiers With Decomposed Vector Rotation-Based Recurrent Neural Networks},
  journal   = {IEEE Trans. Microw. Theory Techn.},
  volume    = {70},
  number    = {11},
  pages     = {4900--4909},
  month     = nov,
  year      = {2022}
}

@article{Hochreiter1997LSTM,
  author    = {Hochreiter, Sepp and Schmidhuber, J{\"u}rgen},
  title     = {Long Short-Term Memory},
  journal   = {Neural Comput.},
  volume    = {9},
  number    = {8},
  month     = nov,
  year      = {1997},
  pages     = {1735--1780},
  doi       = {10.1162/neco.1997.9.8.1735}
}

@inproceedings{Cho2014GRU,
  author    = {Cho, Kyunghyun and van Merri{\"e}nboer, Bart and G{\"u}l{\c c}ehre, {\c C}a{\u g}lar and Bahdanau, Dzmitry and Bougares, Fethi and Schwenk, Holger and Bengio, Yoshua},
  month     = oct,
  title     = {Learning Phrase Representations Using {RNN} Encoder--Decoder for Statistical Machine Translation},
  booktitle = {Proc. Conf. Empirical Methods Natural Lang. Process. (EMNLP)},
  year      = {2014},
  pages     = {1724--1734}
}

@article{Hu2022RVTDCNN,
  author    = {Hu, Xin and Liu, Zhijun and Yu, Xiaofei and Zhao, Yulong and Chen, Wenhua and Hu, Biao and Du, Xuekun and Li, Xiang and Helaoui, Mohamed and Wang, Weidong and Ghannouchi, Fadhel M.},
  journal   = {IEEE Trans. Neural Netw. Learn. Syst.},
  title     = {Convolutional Neural Network for Behavioral Modeling and Predistortion of Wideband Power Amplifiers},
  year      = {2022},
  volume    = {33},
  number    = {8},
  pages     = {3923--3937},
  doi       = {10.1109/TNNLS.2021.3054867}
}

@inproceedings{wu2024opendpd,
  title     = {{OpenDPD}: An Open-Source End-to-End Learning \& Benchmarking Framework for Wideband Power Amplifier Modeling and Digital Pre-Distortion},
  author    = {Wu, Yizhuo and Singh, Gagan Deep and Beikmirza, Mohammadreza and de Vreede, Leo C. N. and Alavi, Morteza and Gao, Chang},
  booktitle = {Proc. IEEE Int. Symp. Circuits Syst. (ISCAS)},
  pages     = {1--5},
  year      = {2024},
  doi       = {10.1109/ISCAS58744.2024.10558162}
}

@article{10994208,
  author    = {Li, Wantao and Criado, Ra{\'u}l and Thompson, William and Montoro, Gabriel and Chuang, Kevin and Gilabert, Pere L.},
  journal   = {IEEE J. Microw.},
  title     = {{GPU}-Based Implementation of Pruned Artificial Neural Networks for Digital Predistortion Linearization of Wideband Power Amplifiers},
  year      = {2025},
  volume    = {5},
  number    = {3},
  pages     = {726--738},
  doi       = {10.1109/JMW.2025.3560420}
}

@inproceedings{11449061,
  author    = {Versluis, Manno and Wu, Yizhuo and Gao, Chang},
  booktitle = {Proc. 35th Int. Conf. Field-Program. Logic Appl. (FPL)},
  title     = {{SparseDPD}: A Sparse Neural Network-Based Digital Predistortion {FPGA} Accelerator for {RF} Power Amplifier Linearization},
  year      = {2025},
  pages     = {154--158},
  doi       = {10.1109/FPL68686.2025.00031}
}

@inproceedings{Duan2025TCN-DPD,
  title     = {{TCN-DPD}: Parameter-Efficient Temporal Convolutional Networks for Wideband Digital Predistortion},
  author    = {Huanqiang Duan and Manno Versluis and Qinyu Chen and Leo C. N. de Vreede and Chang Gao},
  year      = {2025},
  booktitle = {Proc. IEEE/MTT-S Int. Microw. Symp. (IMS)},
  pages     = {1103--1106},
  doi       = {10.1109/IMS40360.2025.11103923}
}

@article{Huang2020,
  author    = {Huang, Hai and Xia, Jingjing and Boumaiza, Slim},
  journal   = {IEEE Trans. Microw. Theory Techn.},
  title     = {Novel Parallel-Processing-Based Hardware Implementation of Baseband Digital Predistorters for Linearizing Wideband {5G} Transmitters},
  year      = {2020},
  volume    = {68},
  number    = {9},
  pages     = {4066--4076},
  doi       = {10.1109/TMTT.2020.2993236}
}

@article{Yue2022,
  author    = {Li, Yue and Wang, Xiaoyu and Zhu, Anding},
  journal   = {IEEE Trans. Microw. Theory Techn.},
  title     = {Reducing Power Consumption of Digital Predistortion for {RF} Power Amplifiers Using Real-Time Model Switching},
  year      = {2022},
  volume    = {70},
  number    = {3},
  pages     = {1500--1508},
  doi       = {10.1109/TMTT.2021.3132347}
}

@inproceedings{Cappello2022,
  author    = {Cappello, T. and Jindal, G. and Nunez-Yanez, J. and Morris, K.},
  title     = {Power Consumption and Linearization Performance of a Bit- and Frequency-Scalable {AM/AM} {AM/PM} Pre-Distortion on {FPGA}},
  booktitle = {Proc. Int. Workshop Integr. Nonlinear Microw. Millimetre-Wave Circuits (INMMiC)},
  pages     = {1--3},
  year      = {2022},
  doi       = {10.1109/INMMiC54248.2022.9762118}
}

@article{Rawat2010,
  author    = {Rawat, M. and Rawat, K. and Ghannouchi, F. M.},
  title     = {Adaptive Digital Predistortion of Wireless Power Amplifiers/Transmitters Using Dynamic Real-Valued Focused Time-Delay Line Neural Networks},
  journal   = {IEEE Trans. Microw. Theory Techn.},
  volume    = {58},
  number    = {1},
  pages     = {95--104},
  month     = jan,
  year      = {2010},
  doi       = {10.1109/TMTT.2009.2036334}
}

@inproceedings{Molendijk2023,
  author    = {Molendijk, Maarten J. and de Putter, Floran A. M. and Gomony, Manil Dev and J{\"a}{\"a}skel{\"a}inen, Pekka and Corporaal, Henk},
  booktitle = {Proc. IEEE 41st Int. Conf. Comput. Design (ICCD)},
  title     = {{BrainTTA}: A 28.6 {TOPS/W} Compiler Programmable Transport-Triggered {NN} {SoC}},
  year      = {2023},
  pages     = {78--85},
  doi       = {10.1109/ICCD58817.2023.00022}
}

@inproceedings{Lin2020,
  author    = {Lin, Chien-Hung and Cheng, Chih-Chung and Tsai, Yi-Min and Hung, Sheng-Je and Kuo, Yu-Ting and Wang, Perry H. and Tsung, Pei-Kuei and Hsu, Jeng-Yun and Lai, Wei-Chih and Liu, Chia-Hung and Wang, Shao-Yu and Kuo, Chin-Hua and Chang, Chih-Yu and Lee, Ming-Hsien and Lin, Tsung-Yao and Chen, Chih-Cheng},
  booktitle = {IEEE Int. Solid-State Circuits Conf. (ISSCC) Dig. Tech. Papers},
  title     = {7.1 {A} 3.4-to-13.3{TOPS/W} 3.6{TOPS} Dual-Core Deep-Learning Accelerator for Versatile {AI} Applications in 7nm {5G} Smartphone {SoC}},
  year      = {2020},
  pages     = {134--136},
  doi       = {10.1109/ISSCC19947.2020.9063111}
}

@article{gao2020edgedrnn,
  author    = {Gao, Chang and Rios-Navarro, Antonio and Chen, Xi and Liu, Shih-Chii and Delbruck, Tobi},
  journal   = {IEEE J. Emerg. Sel. Topics Circuits Syst.},
  title     = {{EdgeDRNN}: Recurrent Neural Network Accelerator for Edge Inference},
  year      = {2020},
  volume    = {10},
  number    = {4},
  pages     = {419--432},
  doi       = {10.1109/JETCAS.2020.3040300}
}

@article{gao2024spartus,
  author    = {Gao, Chang and Delbruck, Tobi and Liu, Shih-Chii},
  journal   = {IEEE Trans. Neural Netw. Learn. Syst.},
  title     = {Spartus: A 9.4 {TOp/s} {FPGA}-Based {LSTM} Accelerator Exploiting Spatio-Temporal Sparsity},
  year      = {2024},
  volume    = {35},
  number    = {1},
  pages     = {1098--1112},
  doi       = {10.1109/TNNLS.2022.3180209}
}

@inproceedings{Li2025ISCAS,
  author    = {Li, Ang and Wu, Haolin and Wu, Yizhuo and Chen, Qinyu and de Vreede, Leo C. N. and Gao, Chang},
  booktitle = {Proc. IEEE Int. Symp. Circuits Syst. (ISCAS)},
  title     = {{DPD-NeuralEngine}: A 22-nm 6.6-{TOPS/W}/mm$^2$ Recurrent Neural Network Accelerator for Wideband Power Amplifier Digital Pre-Distortion},
  year      = {2025},
  pages     = {1--5},
  doi       = {10.1109/ISCAS56072.2025.11043563}
}

@article{cao2017modified,
  title     = {A Modified Decomposed Vector Rotation-Based Behavioral Model With Efficient Hardware Implementation for Digital Predistortion of {RF} Power Amplifiers},
  author    = {Cao, Wenhui and Zhu, Anding},
  journal   = {IEEE Trans. Microw. Theory Techn.},
  volume    = {65},
  number    = {7},
  pages     = {2443--2452},
  year      = {2017}
}

@article{Othmani2023ParallelDSM,
  author    = {Othmani, Marouan and Boulejfen, Noureddine and Turunen, Matias and All{\'e}n, Markus and Ghannouchi, Fadhel M. and Valkama, Mikko},
  journal   = {IEEE Trans. Circuits Syst. I, Reg. Papers},
  title     = {Parallel Delta-Sigma Modulator-Based Digital Predistortion of Wideband {RF} Power Amplifiers},
  year      = {2023},
  volume    = {70},
  number    = {2},
  pages     = {705--718},
  doi       = {10.1109/TCSI.2022.3218295}
}

@article{11278888,
  author    = {Kim, Daegyeom and Park, Sehoon and Kim, Sunmean},
  journal   = {IEEE Trans. Circuits Syst. II, Exp. Briefs},
  title     = {{LW-ResGRU} {DPD}: Layer-Wise Residual {GRU} for Low-Latency {DPD} in Nonlinear {RF} Systems},
  year      = {2026},
  volume    = {73},
  number    = {2},
  pages     = {153--157},
  doi       = {10.1109/TCSII.2025.3640703}
}

@article{11373642,
  author    = {Khan, Mohd Tasleem and Ding, Yuan and Goussetis, George},
  journal   = {IEEE Trans. Neural Netw. Learn. Syst.},
  title     = {Next-Gen Digital Predistortion From Hardware Acceleration of Neural Networks: Trends, Challenges, and Future},
  year      = {2026},
  volume    = {37},
  number    = {8},
  pages     = {3503--3523},
  doi       = {10.1109/TNNLS.2026.3656642}
}

@article{PhaseNormalizedNN2023,
  author    = {Fischer-B{\"u}hner, Arne and Anttila, Lauri and Gomony, Manil Dev and Valkama, Mikko},
  journal   = {IEEE Microw. Wireless Technol. Lett.},
  title     = {Phase-Normalized Neural Network for Linearization of {RF} Power Amplifiers},
  year      = {2023},
  volume    = {33},
  number    = {9},
  pages     = {1357--1360},
  month     = sep,
  doi       = {10.1109/LMWT.2023.3290980}
}

@article{PNRNN,
  author    = {Fischer-B{\"u}hner, Arne and Anttila, Lauri and Gomony, Manil Dev and Valkama, Mikko},
  journal   = {IEEE Microw. Wireless Technol. Lett.},
  title     = {Recursive Neural Network With Phase-Normalization for Modeling and Linearization of {RF} Power Amplifiers},
  year      = {2024},
  volume    = {34},
  number    = {6},
  pages     = {809--812},
  doi       = {10.1109/LMWT.2024.3393859}
}

@misc{AD9375,
  author       = {{Analog Devices}},
  title        = {{AD9375}: Integrated, Dual {RF} Transceiver With Observation Path, Data Sheet Rev.~0},
  year         = {2017},
  howpublished = {\url{https://www.analog.com/media/en/technical-documentation/data-sheets/ad9375.pdf}}
}

@electronic{AD9375UG992,
  author       = {{Analog Devices}},
  organization = {Analog Devices, Inc.},
  title        = {{AD9371}/{AD9375} System Development User Guide ({UG-992}, Rev.~{B})},
  note         = {{DPD} overview, p.~319},
  url          = {https://www.analog.com/media/en/technical-documentation/user-guides/ug-992.pdf}
}

@techreport{3gpp38104,
  author      = {{3rd Generation Partnership Project}},
  title       = {{NR}; Base Station ({BS}) Radio Transmission and Reception},
  institution = {ETSI},
  type        = {Technical Specification},
  number      = {TS 38.104, V18.8.0},
  year        = {2025},
  month       = jan,
  url         = {https://www.etsi.org/deliver/etsi_ts/138100_138199/138104/18.08.00_60/ts_138104v180800p.pdf}
}

@article{BOJANET,
  author    = {Zhang, Qianqian and Jiang, Chengye and Yang, Guichen and Han, Renlong and Liu, Falin},
  journal   = {IEEE Trans. Microw. Theory Techn.},
  title     = {Block-Oriented Recurrent Neural Network for Digital Predistortion of {RF} Power Amplifiers},
  year      = {2024},
  volume    = {72},
  number    = {7},
  pages     = {3875--3885},
  doi       = {10.1109/TMTT.2023.3337939}
}

@article{APNRRU,
  author    = {Fischer-B{\"u}hner, Arne and Anttila, Lauri and Turunen, Matias and Gomony, Manil Dev and Valkama, Mikko},
  journal   = {IEEE Trans. Microw. Theory Techn.},
  title     = {Augmented Phase-Normalized Recurrent Neural Network for {RF} Power Amplifier Linearization},
  year      = {2025},
  volume    = {73},
  number    = {1},
  pages     = {412--422},
  doi       = {10.1109/TMTT.2024.3484581}
}

@article{Chen2022MANGMP,
  author    = {Chen, Wenhua and Liu, Xin and Chu, Jiaming and Wu, Huibo and Feng, Zhenghe and Ghannouchi, Fadhel M.},
  journal   = {IEEE Trans. Circuits Syst. I, Reg. Papers},
  title     = {A Low Complexity Moving Average Nested {GMP} Model for Digital Predistortion of Broadband Power Amplifiers},
  year      = {2022},
  volume    = {69},
  number    = {5},
  pages     = {2070--2083},
  doi       = {10.1109/TCSI.2022.3150408}
}

@article{Wu2024DCCDPD,
  author    = {Wu, Xiaofang and Pan, Xiao and Sun, Peng and Zhou, Jianyang},
  journal   = {IEEE Trans. Circuits Syst. I, Reg. Papers},
  title     = {Scalable Double Crosstalk Canceling Digital Predistortion for {MIMO} Transmitters},
  year      = {2024},
  volume    = {71},
  number    = {12},
  pages     = {6292--6305},
  doi       = {10.1109/TCSI.2024.3439580}
}

@misc{van2018unreasonable,
  author       = {van der Westhuizen, Jos and Lasenby, Joan},
  title        = {The Unreasonable Effectiveness of the Forget Gate},
  year         = {2018},
  howpublished = {arXiv:1804.04849},
  url          = {https://arxiv.org/abs/1804.04849}
}

@misc{OpenDPDSoftware,
  author       = {{OpenDPD Contributors}},
  title        = {{OpenDPD}: Software Repository},
  howpublished = {\url{https://github.com/lab-emi/OpenDPD}},
  note         = {Accessed: Nov.~13, 2026}
}

@article{10502240,
  author    = {Wu, Yizhuo and Li, Ang and Beikmirza, Mohammadreza and Singh, Gagan Deep and Chen, Qinyu and de Vreede, Leo C. N. and Alavi, Morteza and Gao, Chang},
  journal   = {IEEE Microwave and Wireless Technology Letters},
  title     = {{MP-DPD}: Low-Complexity Mixed-Precision Neural Networks for Energy-Efficient Digital Predistortion of Wideband Power Amplifiers},
  year      = {2024},
  volume    = {34},
  number    = {6},
  pages     = {817--820},
  doi       = {10.1109/LMWT.2024.3386330}
}

@misc{Wu2025OpenDPDv2,
  author       = {Wu, Yizhuo and Li, Ang and Gao, Chang},
  title        = {{OpenDPDv2}: A Unified Learning and Optimization Framework for Neural Network Digital Predistortion},
  year         = {2025},
  howpublished = {arXiv:2507.06849},
  url          = {https://arxiv.org/abs/2507.06849}
}

@article{Zhang1996Sigmoid,
  author    = {Zhang, Ming and Vassiliadis, Stamatis and Delgado-Frias, Jos{\'e} G.},
  title     = {Sigmoid Generators for Neural Computing Using Piecewise Approximations},
  journal   = {IEEE Trans. Comput.},
  volume    = {45},
  number    = {9},
  pages     = {1045--1049},
  year      = {1996},
  month     = sep,
  doi       = {10.1109/12.537127}
}

@inproceedings{Howard2019MobileNetV3,
  author    = {Howard, Andrew and Sandler, Mark and Chu, Grace and Chen, Liang-Chieh and Chen, Bo and Tan, Mingxing and Wang, Weijun and Zhu, Yukun and Pang, Ruoming and Vasudevan, Vijay and Le, Quoc V. and Adam, Hartwig},
  title     = {Searching for {MobileNetV3}},
  booktitle = {Proc. IEEE/CVF Int. Conf. Comput. Vis. (ICCV)},
  pages     = {1314--1324},
  year      = {2019},
  doi       = {10.1109/ICCV.2019.00140}
}

@techreport{Breinbauer2024EHTEVM,
  author      = {Breinbauer, M.},
  title       = {Hints for {IEEE} 802.11be {EVM} Measurements},
  institution = {Rohde \& Schwarz},
  type        = {Application Note},
  number      = {1EF114, Version 1e},
  year        = {2024},
  month       = aug,
  url         = {https://scdn.rohde-schwarz.com/ur/pws/dl_downloads/dl_application/application_notes/1ef114/1EF114_1e_802_11be_EVM.pdf}
}

\end{document}